\documentclass[twocolumn,10pt]{asme2ej}
\usepackage{color}
\usepackage{graphicx} 
\usepackage{bm}
\usepackage{amsmath}
\usepackage{amssymb}
\usepackage{multirow}
\usepackage[normalem]{ulem}
\title{Fluid Mechanics of Liquid Metal Batteries}

\author{Douglas H. Kelley
  \affiliation{
    Department of Mechanical Engineering\\
    University of Rochester\\
    Rochester, NY 14627\\
    Email: d.h.kelley@rochester.edu
  }	
}

\author{Tom Weier
  \affiliation{Institute of Fluid Dynamics\\
    Helmholtz-Zentrum Dresden - Rossendorf\\
    Bautzner Landstr. 400\\
    01328 Dresden, Germany\\
    Email: t.weier@hzdr.de
  }
}

\begin{document}
\maketitle    

\begin{abstract}
\emph{ 
The design and performance of liquid metal batteries, a new technology
for grid-scale energy storage, depend on fluid mechanics because the
battery electrodes and electrolytes are entirely liquid. Here we
review prior and current research on the fluid mechanics of liquid
metal batteries, pointing out opportunities for future
studies. Because the technology in its present form is just a few
years old, only a small number of publications have so far considered
liquid metal batteries specifically. We hope to encourage
collaboration and conversation by referencing as many of those publications as possible here. Much can also be learned by linking to extensive prior literature considering phenomena observed or expected in liquid metal batteries, including thermal convection, magnetoconvection, Marangoni flow, interface instabilities, the Tayler instability, and electro-vortex flow. We focus on phenomena, materials, length scales, and current densities relevant to the liquid metal battery designs currently being commercialized. We try to point out breakthroughs that could lead to design improvements or make new mechanisms important. 
} 
\end{abstract}

The story of fluid mechanics research in liquid metal batteries (LMBs) begins with one very important application: grid-scale storage. Electrical grids have almost no energy storage capacity, and adding storage will make them more robust and more resilient even as they incorporate increasing amounts of intermittent and unpredictable wind and solar generation. Liquid metal batteries have unique advantages as a grid-scale storage technology, but their uniqueness also means that designers must consider chemical and physical mechanisms~--- including fluid mechanisms~--- that are relevant to few other battery technologies, and in many cases not yet well-understood. We will review the fluid mechanics of liquid metal batteries, focusing on studies undertaken with that technology in mind, and also drawing extensively from prior work considering similar mechanisms in other contexts. In the interest of promoting dialogue across this new field, we have endeavored to include the work of many different researchers, though inevitably some will have eluded our search, and we ask for the reader's sympathy for regrettable omissions. Our story will be guided by technological application, focusing on mechanisms most relevant to liquid metal batteries as built for grid-scale storage. We will consider electrochemistry and theoretical fluid mechanics only briefly because excellent reviews of both topics are already available in the literature. In \S\ref{sec:intro} below we provide an overview and brief introduction to liquid metal batteries, motivated by the present state of worldwide electrical grids, including the various types of liquid metal batteries that have been developed. We consider the history of liquid metal batteries in more detail in \S\ref{sec:history}, connecting to the thermally regenerative electrochemical cells developed in the middle of the twentieth century. Continuing, we consider the fluid mechanisms that are most relevant to liquid metal batteries: thermal convection and magnetoconvection in \S\ref{sec:convection}, Marangoni flow in \S\ref{sec:Marangoni}, interface instabilities in \S\ref{sec:interface}, the Tayler instability in \S\ref{sec:Tayler}, and electro-vortex flow in \S\ref{sec:electro-vortex}. We conclude with a summary and reflection on future directions in \S\ref{sec:summary}. 

\section{Introduction}
\label{sec:intro}


A typical electrical grid spans a country or a continent, serving
millions of consumers by linking them to an intricate network of
hundreds or thousands of large generators. A grid can be understood as
a single, gigantic machine, because all of its rotating generators
must spin in synchrony, and within a fraction of a percent of their
design speed, in order for the grid to function properly. Changes to
any one part of the grid affect all parts of the grid. The
implications of this interconnectedness are made more profound by the
fact that today's grids have nearly zero storage capacity. When more
electricity is being consumed than generated, conservation of energy
requires that the kinetic energy of the spinning generators drop, so
they slow down, quickly losing synchrony, damaging equipment, and
causing brownouts or blackouts if left unchecked. Conversely, when
more electricity is being generated than consumed, generators speed
up, risking all the same problems. Fluctuations in demand are as old
as electrical utilities, and have historically been managed by
continually adjusting supply by turning generators on and off. Now,
grids must also accommodate fluctuations in supply, as intermittent
wind and solar generation expand rapidly because of their plummeting
costs and the long-term imperative that humankind generate a
significant share of our energy using renewable
sources~\cite{Kassakian:2011}. Large-scale storage on electrical grids
would enable widespread deployment of renewable
generation~\cite{Whittingham:2012,Backhaus:2013} while maintaining
stability~\cite{NardelliRubidoWangEtAl:2014}. Many technologies for
grid-scale storage have been proposed, including pumped hydro (which
accounts for the vast majority of existing storage), pressurized air,
thermal storage, flywheels, power-to-gas and batteries. Liquid metal
batteries are a particular grid-scale storage technology that comes
with interesting fluid mechanical challenges.
  
Like any battery, a liquid metal battery discharges by allowing an energetically-favorable chemical reaction to proceed in a controlled way. Control is maintained by separating the two reactants (the electrodes) with an electrolyte that prevents electrode materials from passing if they are neutral, but allows them to pass if they are ionized. Thus the reaction proceeds only if some other path passes matching electrons, which then recombine with the ions and go on to react. The other path is the external circuit where useful work is done, thanks to the energy of the flowing electrons. The battery can later be recharged by driving electrons in the opposite direction, so that matching ions come along as well. 

Battery electrodes can be made from a wide variety of materials, including liquid metals. For example, a liquid sodium negative electrode (anode) can be paired with a sulfur positive electrode (cathode) and a solid $\beta$-alumina electrolyte. (Here and throughout, we assign the names ``anode'' and ``cathode'' according to the roles played during discharge.) Na$||$S batteries operate at about 300\,$\mbox{}^\circ$C and have been deployed for grid-scale storage. ZEBRA batteries~\cite{BonesTeagleBrookerCullen:1989,Sudworth:2001}, named for the Zeolite Battery Research Africa Project that developed them, use a NaAlCl$_4$ negative electrode that allows them to operate at temperatures as low as 245\,$\mbox{}^\circ$C. An electrolyte composed of Na-doped $\beta$-alumina conducts Na\textsuperscript{+} ions. Lower operating temperatures are possible with batteries in which a Na negative electrode is combined with a NiCl$_2$ positive electrode and a NaAlCl$_4$ electrolyte, separated from the negative electrode with $\beta$-alumina to prevent corrosion. Alloying Cs with the Na can substantially improve wetting to $\beta$-alumina, allowing battery operation at still lower temperatures~\cite{Lu:2014}. Sumitomo has recently documented a battery design using a eutectic mix of potassium and sodium bis(fluorosupfonyl)amide salts along with electrodes made from unspecified sodium compounds~\cite{Fukunaga:2012,Nitta:2013}. These battery designs and others like them involve liquid metals but require a solid separator between the layers. 

\begin{figure}
\centering
\includegraphics[width=\columnwidth]{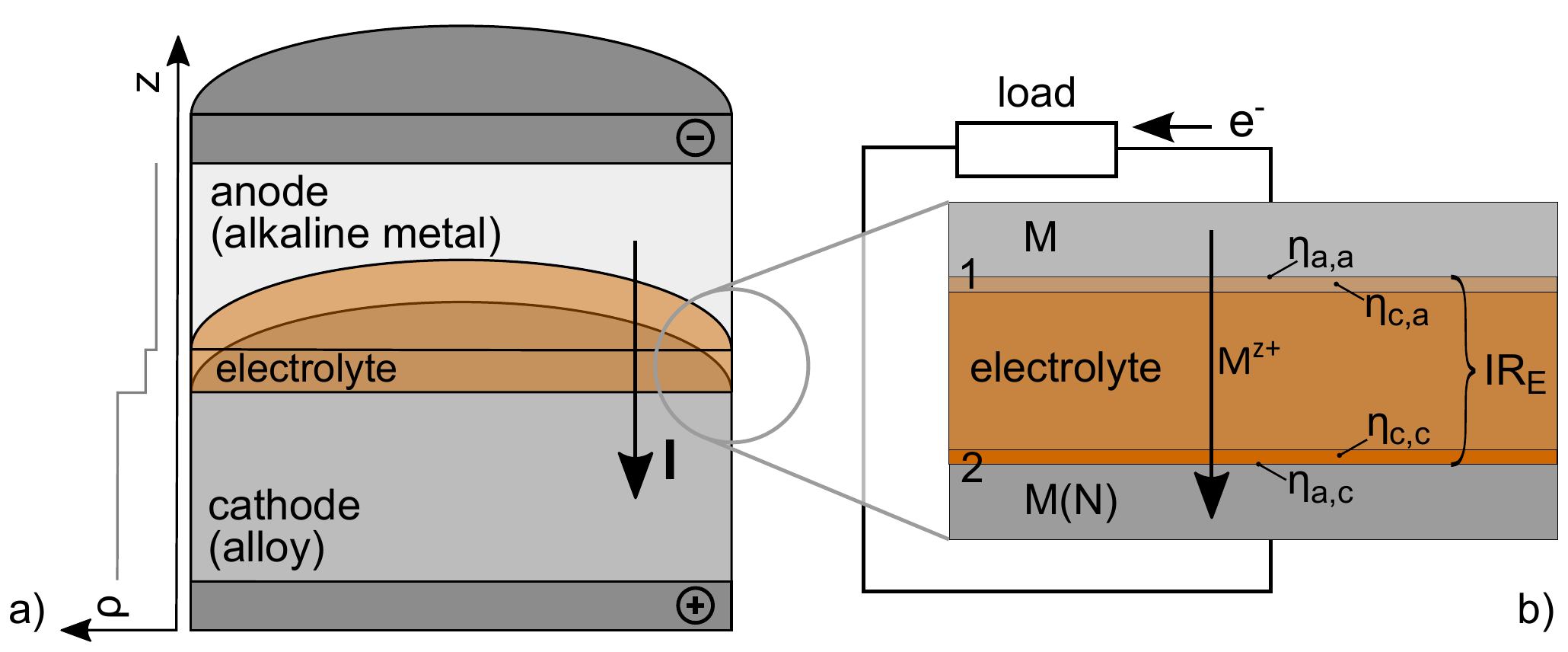}
\caption{\label{fig:lmb_sketch}Sketch of a liquid metal cell with
  discharge current and density profile for fully charged state and
  isothermal conditions (a) and schematic discharge process (b) from \cite{WeierBundEl-MofidHorstmannLalauLandgrafNimtzStaraceStefaniWeber:2017}.}
\end{figure}

\begin{figure}
  \centering
  \includegraphics[width=\columnwidth]{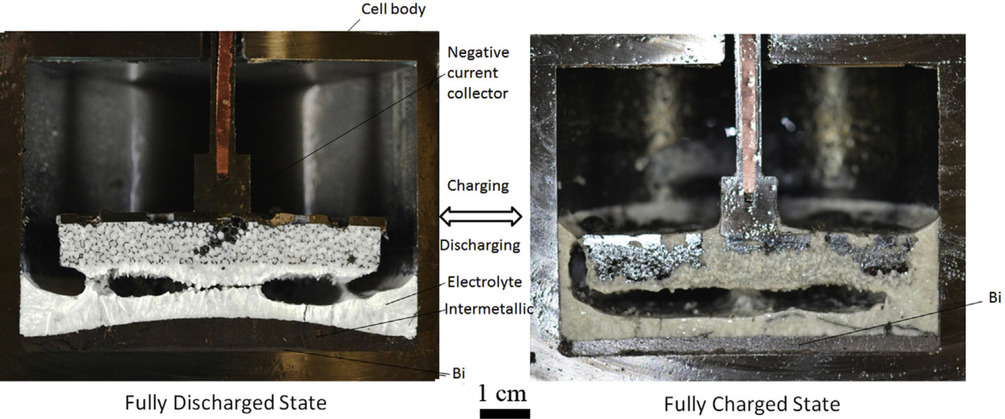}
  \caption{\label{fig:Ning2015fig5b}Cross-sections of prototype liquid metal batteries. Both are enclosed in a stainless steel casing that also serves as the positive current collector, and both have a foam negative current collector attached to a copper conductor that exits the top of the battery. In the discharged state (left), the foam is nearly filled with electrolyte, and a dark Li-Bi intermetallic layer is visible at bottom. In the charged state (right), lithium metal is visible in the foam, and the positive electrode at bottom has been restored to nearly pure bismuth. Because these photographs were taken at room temperature, the electrolyte does not fill the volume between the electrodes, but during operation, it would. The space above the negative current collector is filled with inert gas during operation. Adapted from~\cite{Ning:2015}, with permission.}
  \end{figure}

As discovered at Argonne National Laboratory in the 1960s~\cite{CairnsCrouthamelFischerEtAl:1967} and rediscovered at MIT recently~\cite{KimBoysenNewhouseEtAl:2013}, batteries can also be designed with liquid metal electrodes and molten salt electrolytes, requiring no separator at all. We shall use the term ``liquid metal batteries'' to refer to those designs specifically. An example is sketched in Fig.~\ref{fig:lmb_sketch}a, and cross-sections of two laboratory prototypes are shown in Fig.~\ref{fig:Ning2015fig5b}. The internal structure of the battery is maintained by gravity, since negative electrode materials typically have lower density than electrolyte materials, which have lower density than positive electrode materials. A solid metal positive current collector contacts the positive electrode, and usually serves as the container as well. A solid metal negative current collector connects to the negative electrode and is electrically insulated from the positive current collector. 

Because the negative electrode is liquid and the positive current collector is also the battery vessel, some care is required to prevent shorts between them. It is possible to electrically insulate the positive current collector by lining it with a ceramic, but ceramic sleeves are too expensive for grid-scale applications and are prone to cracking. Instead, typical designs separate the liquid metal negative electrode from vessel walls with a metal foam, as shown in Fig.~\ref{fig:fac_sketch}. The high surface tension of the liquid metal provides sufficient capillary forces to keep it contained in the pores of the foam. The solid foam also inhibits flow in the negative electrode, which is likely negligible at length scales larger than the pore size. In many designs, the foam is held in place by a rigid conductor, as shown, so that its height stays constant. However, as the battery discharges and the positive electrode becomes a pool of two-part alloy, it swells. If the positive electrode swells enough to contact the foam, a short occurs, so the foam height must be carefully chosen, taking into account the thickness of the positive electrode and the density change it will undergo during discharge. 

Most liquid metal cells are concentration cells. Their open circuit
voltage (OCV) is solely given by the activity of the alkaline metal in
the cathode alloy. The equations for the transfer reactions at the two
interfaces (see Fig.~\ref{fig:lmb_sketch}b) read:
\begin{eqnarray}
    \text{M} & \rightarrow & \text{M}^{z+} + z \text{e}^{-}\\
  \text{M}^{z+} + z \text{e}^{-} & \rightarrow & \text{M(N)}
\end{eqnarray}
for the anode/electrolyte interface (1) and the electrolyte/cathode
interface (2) during discharge. M denotes an alkali ($z=1$) or
earth-alkali ($z=2$) metal of the negative electrode, and N refers to the
heavy or half metal of the positive electrode. A variety of chemistries have been demonstrated, including Mg$||$Sb~\cite{BradwellKimSirkSadoway:2012}, Li$||$Pb-Sb~\cite{WangJiangChungOuchiBurkeBoysenKimMueckeSadoway:2014}, Li$||$Bi~\cite{Ning:2015}, Na$|$NaCl-CaCl$_2|$Zn~\cite{XuKjosOsenEtAl:2016, XuMartinezOsenKjosKongsteinHaarberg:2017} and Ca-Mg$||$Bi~\cite{KimBoysenOuchiEtAl:2013,Ouchi:2016}. (See~\cite{LiYinWangEtAl:2016} for a review.) The Li$||$Pb-Sb chemistry has been studied most, and is typically paired with a triple-eutectic LiF-LiCl-LiI electrolyte because of its relatively low melting temperature (about 341\,$\mbox{}^\circ$C \cite{CairnsCrouthamelFischerEtAl:1967,JohnsonHathaway:1971}). The equilibrium
potentials $\varphi_{0}$ of both half-cells can be written as
\begin{eqnarray}
  \varphi_{0}(1) &=& \varphi_{00} + \frac{R T}{z F} \ln{\frac{a_{\text{M}^{z+}}}{a_{\text{M}}}}\\
  \varphi_{0}(2) &=& \varphi_{00} + \frac{R T}{z F} \ln{\frac{a_{\text{M}^{z+}}}{a_{\text{M(N)}}}}
\end{eqnarray}
with the standard potential $\varphi_{00}$, the universal gas constant $R$, the temperature $T$, the Faraday constant $F$ and the activity $a$ of the metal in the pure (M), ionic (M$^{z+}$) and the alloyed (M(N)) state. The difference of the two electrode potentials $\varphi_0(2)$ and $\varphi_0(1)$ is the cell's OCV 
\begin{equation}
E_{\text{OC}} = - \frac{R T}{z F} \ln{a_{\text{M(N)}}}.
\end{equation}
Only the activity of the alkali metal in the alloy $a_{\text{M(N)}}$
determines the OCV since the standard potentials of both half cells
are identical and the activity of the pure anode is one by definition.
Under current flow, only the terminal voltage $E$ is available. It is
the difference of OCV and several terms describing voltage losses,
i.e., polarizations (cf. \cite{Swinkels:1971} and
Fig.~\ref{fig:lmb_sketch}b) occurring under current ($I$) flow:
\begin{equation}
E = E_{\text{OC}} - IR_{E} - \eta_{c, a} - \eta_{c, c} - \eta_{a, a} - \eta_{a, c}.
\label{eq:terminal_voltage}
\end{equation}
These voltage losses are due to the electrolyte resistance
$R_{\text{E}}$, the concentration polarizations at the anode
$\eta_{c, a}$ and cathode $\eta_{c, c}$ and the corresponding
activation potentials $\eta_{a, a}$ and $\eta_{a, c}$.  Typically,
ohmic losses dominate activation and concentration polarizations by
far, but mass transfer limitations may nevertheless sometimes occur in
the cathodic alloy.

Liquid metal batteries have advantages for grid-scale
storage. Eliminating solid separators reduces cost and eliminates the possibility of failure from a cracked separator. Perhaps more importantly, solid
separators typically allow much slower mass transport than liquids, so
eliminating solids allows faster charge and discharge with smaller
voltage losses. Liquid electrodes improve battery life, because the
life of Li-ion and other more traditional batteries is limited when
their solid electrodes are destroyed due to repeated shrinking and
swelling during charge and discharge. Projections from experimental
measurements predict Li$||$Pb-Sb batteries will retain 85\% of their
capacity after daily discharge for 
ten years~\cite{WangJiangChungOuchiBurkeBoysenKimMueckeSadoway:2014}. The
Li$||$Pb-Sb chemistry is composed of Earth-abundant elements available
in quantities large enough to provide many GWh of storage. Low cost is
also critical if a technology is to be deployed
widely~\cite{SpatoccoSadoway:2015}, and liquid metal batteries are
forecast to have costs near the \$100/kWh target set by
the US Advanced Research Projects Agency-Energy (ARPA-e). Their energy and power density are moderate, and substantially below the Li-ion batteries that are ubiquitous in portable electronics, but density is less essential than cost in stationary grid-scale storage. Li-ion batteries today cost substantially more than \$100/kWh, but their costs have dropped continually over time and will likely drop substantially more as the Tesla GigaFactory 1, the world's largest Li-ion battery plant, continues to increase its production. The energy efficiency of liquid metal batteries varies widely with current density, but at a typical design value of 275~mA/cm\textsuperscript{2} is 73\%~\cite{WangJiangChungOuchiBurkeBoysenKimMueckeSadoway:2014}, similar to pumped hydro storage.

Liquid metal batteries also present challenges. During discharge, Li$||$Pb-Sb batteries provide only about 0.8~V~\cite{WangJiangChungOuchiBurkeBoysenKimMueckeSadoway:2014}. Despite variation with battery chemistry, all conventional liquid metal batteries have voltage significantly less than Li-ion batteries. Lacking solid separators, liquid metal batteries are not suitable for portable applications in which disturbing the fluid layers could rupture the electrolyte layer, causing electrical shorts between the positive and negative electrodes and destroying the battery. Rupture might also result from vigorous fluid flows even if the battery is stationary, such as the Tayler instability (\S\ref{sec:Tayler}), interface instabilities (\S\ref{sec:interface}), Marangoni flow (\S\ref{sec:Marangoni}), electro-vortex flow (\S\ref{sec:electro-vortex}), or their combination. Flow mechanisms may also interact, triggering instabilities more readily. Existing liquid metal battery chemistries require high operating temperatures (475\,$\mbox{}^\circ$C for Li$||$Pb-Sb). Little energy is wasted heating large batteries because Joule heating (losses to electrical resistance) provides more than enough energy to maintain the temperature. Still, high temperatures promote corrosion and make air-tight mechanical seals difficult. Finally, poor mixing during discharge can cause local regions of a liquid metal electrode to form unintended intermetallic solids that can eventually span from the positive to the negative electrode, destroying the battery. Solid formation may well be the leading cause of failure in liquid metal batteries. 

\section{History and Past Work}
\label{sec:history}

\subsection{Three-layer Aluminium refinement cells}
\label{sec:history-refinement-cells}
\begin{figure*}
  \includegraphics[width=\textwidth]{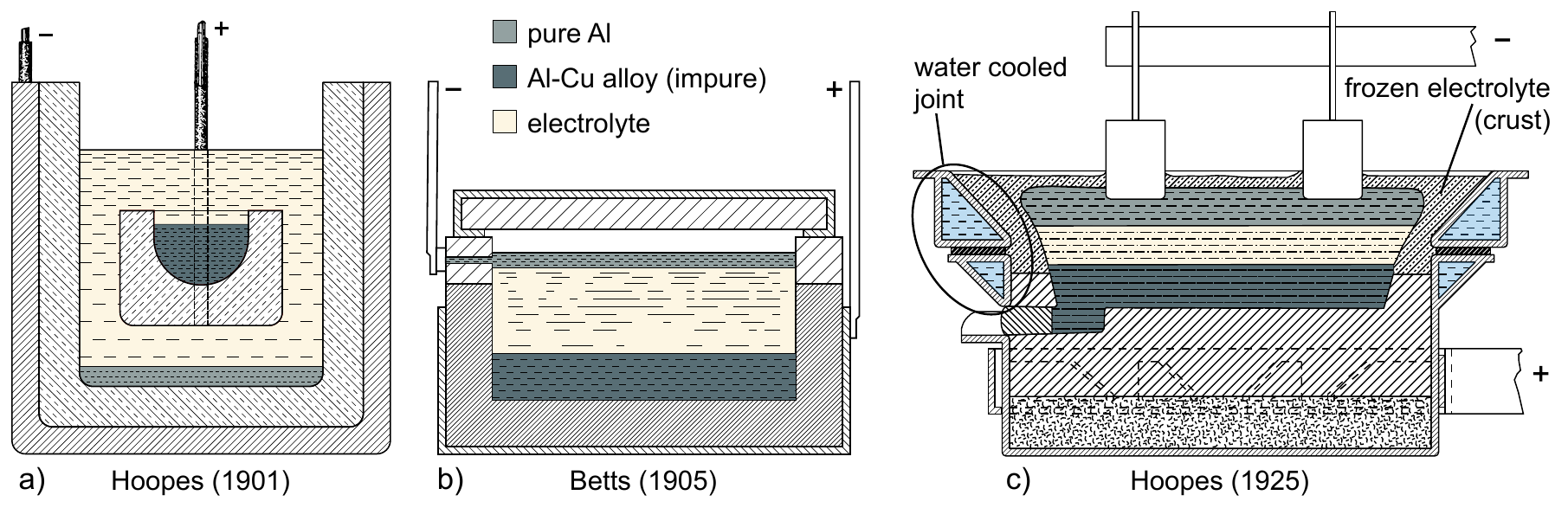}
  \caption{\label{fig:Hoopes_Betts_Hoopes}Aluminium refinement cells adapted from Hoopes (1901, a), Betts (1905, b), and Hoopes (1925, c).}
\end{figure*}
The central idea at the heart of LMBs is the three-layer arrangement
of liquid electrodes and electrolyte. This seemingly simple idea (in
fact so apparently simple that it is sometimes \cite{Drossbach:1952}
questioned if it deserves to be patented at all) did not originate
with LMBs. Instead, using a stable stratification of two liquid metals
interspaced with a molten salt for electrochemical purposes was first
proposed 1905 by Betts \cite{Betts:1905} in the context of aluminium
purification (see Fig.~\ref{fig:Hoopes_Betts_Hoopes}b). However, Betts was not able to commercialize his
process. Instead Hoopes, who had a more complicated arrangement using
a second internal vessel for aluminium electrorefining patented in
1901 \cite{Hoopes:1901} (Fig.~\ref{fig:Hoopes_Betts_Hoopes}a)
developed later a water cooled three-layer cell \cite{Hoopes:1925}
(Fig.~\ref{fig:Hoopes_Betts_Hoopes}c) that could be successfully
operated. According to Frary \cite{Frary:1925} Hoopes as well thought
of using a three-layer cell around 1900. It can be seen from Table~\ref{tab:aluminium} that even if the idea to use three liquid layers
were a trivial one, its realization and transformation to
a working process was highly non-trivial indeed.

The submerged vessel containing the negative electrode, initially suggested by Hoopes~\cite{Hoopes:1901} (Fig.~\ref{fig:Hoopes_Betts_Hoopes}a) is filled
with molten impure aluminium and surrounded by a bath of fused
cryolite. Cryolite is less dense than the pure or impure Al. In the presence of flow, Al dissolves into the cryolite and deposits at the carbon
walls of the outer vessel, and pure Al can be collected at the bottom of
the outer vessel. However, the current density is distributed very
inhomogeneously, concentrating around the opening of the inner
vessel. This implies large energy losses and strong local heating
rendering a stable operation over longer times impossible.

Betts \cite{Betts:1905,Mueller:1932}
(Fig.~\ref{fig:Hoopes_Betts_Hoopes}b) alloyed the impure Al with Cu and
added BaF\textsubscript{2} to the cryolite to increase the density of
the salt mixture and to enable the purified Al to float on the fused
salt. This three-layer arrangement guaranteed the shortest possible current paths
and enabled homogeneous current density distributions. Additionally
the evaporation of the electrolyte was drastically reduced by the Al
top layer. However, under the high operating temperatures the cell
walls became electrically conducting, got covered with metal that
short-circuited the negative and positive electrodes and thus prevented successful
operation of the cell \cite{Mueller:1932}.

Only Hoopes' sophisticated construction
\cite{Hoopes:1925,Frary:1925} (Fig.~\ref{fig:Hoopes_Betts_Hoopes}c)
was finally able to operate for longer times. A key element of Hoopes'
construction is the division of the cell into two electrically
insulated sections. The joint between them is water cooled and thereby
covered by a crust of frozen electrolyte providing electrical as well
as thermal insulation \cite{HoopesEdwardsHorsfield:1925}. A similar
idea was later applied to Na$||$Bi galvanic cells by Shimotake and
Hesson \cite{ShimotakeHesson:1968}.  Additionally, instead of using a
single electric contact to the purified Al at the cells' side as did Betts, Hoopes arranged several graphite current collectors along
the Al surface that provided a more evenly distributed
current. However, the electrolyte used by Hoopes (see
Table~\ref{tab:aluminium}) had a relatively high melting temperature and
a tendency to creep to the surface between the cell walls and the
purified Al \cite{Zeerleder:1955,Frary:1925}.
\nocite{Eger:1955}
According to Beljajew et al.~\cite{BeljajewRapoportFirsanowa:1957}
(see as well Gadeau \cite{Gadeau:1939}), the complicated design of
Hoopes' cell, especially the water cooled walls, prevented continuous
use in production.
It was not until 1934 that super-purity aluminium became widely available 
with Gadeau's \cite{Gadeau:1936} three-layer refining process
that used a different electrolyte (see Table~\ref{tab:aluminium})
according to a patent filed in 1932. Its lower melting point allowed for
considerably decreased operating temperature. Gadeau's cell was lined with
magnesite that could withstand the electrolyte attack without the need
of water cooling. However, the BaCl\textsubscript{2} used in the
electrolyte mixture decomposed partially, so the electrolyte
composition had to be monitored and adjusted during cell operation
when necessary. This difficulty was overcome by using the purely
fluoride based electrolyte composition suggested by Hurter
\cite{Hurter:1937} (see Table~\ref{tab:aluminium}, S.A.I.A. Process).

\begin{table*}
  \caption{\label{tab:aluminium}Characteristics of different three-layer aluminium refining processes (approximate values, adapted from Table~I of \cite{PearsonPhillips:1957} and Table~6 of \cite{BeljajewRapoportFirsanowa:1957})}
  \begin{tabular}{llll}
      \hline
      \hline
    & Hoopes Process & Gadeau Process & S.A.I.A. Process \\
    \hline
    \textbf{top layer} & pure Al & pure Al & pure Al \\
    density / kgm\textsuperscript{-3} & 2290    & 2300 & 2300 \\
    melting point / \mbox{}$^{\circ}$\,C &  660 & 660  & 660 \\
    \textbf{electrolyte} & AlF$_3$-NaF-BaF$_2$ & AlF$_3$-NaF-BaCl$_2$-NaCl & AlF$_3$-NaF-BaF$_2$-CaF$_2$\\
    composition (mass\%)  & 0.34-0.28-0.38 & 0.15-0.17-0.6-0.08 & 0.48-0.18-0.18-0.16\\
    density / kgm\textsuperscript{-3} & 2500 & 2700 & 2500 \\
    melting point / \mbox{}$^{\circ}$\,C & 900 & 700 & 670 \\
    \textbf{bottom layer} & Al-Cu & Al-Cu-Other & Al-Cu \\
    composition (mass\%)  & 0.75-0.25 & 0.6-0.28-0.12 & 0.7-0.3\\
    density / kgm\textsuperscript{-3} & 2800 & 3140 & 3050 \\
    melting point / \mbox{}$^{\circ}$\,C & 550 & unspecified & 590 \\
    \textbf{operating temperature} / \mbox{}$^{\circ}$\,C & 950 & 800 & 750 \\
      \hline
      \hline
  \end{tabular}
\end{table*}

Aluminum refining cells can tolerate larger voltage drops than LMBs, so the electrolyte layer is often much thicker. Values quoted are between 8\,cm
\cite{Dube:1954,BeljajewRapoportFirsanowa:1957} that should be a good estimate for current practice \cite{Wolstenholme:1982}, 10\,cm \cite{Zeerleder:1955}, 20\,cm \cite{YanFray:2010} and 25\,cm
\cite{PearsonPhillips:1957}. These large values are on the one hand
due to the need for heat production. On the other hand a large distance between
the negative and positive electrodes is necessary to prevent flow induced inter-mixing of
the electrode metals that would nullify refinement. It is often
mentioned
\cite{Frary:1925,EdwardsFraryJeffries:1930,PearsonPhillips:1957,Zeerleder:1955} that
strong electromagnetic forces trigger those flows. Unlike aluminium
electrolysis cells, refinement cells have been optimized little, and the technology would certainly gain from new research
\cite{YanFray:2010}. Yan and Fray \cite{YanFray:2010} directly
invoke the low density differences as a cause for the instability of the
interfaces, discussed here in \S\ref{sec:interface}. They
attribute the limited application of fused salt electrorefining to the
present design of refining cells that does not take advantage of the
high electrical conductivity and the very low thermodynamic potential
  required for the process. Coupling optimized electrorefining to
  carbon-free generation of electricity should, according to Yan and
  Fray \cite{YanFray:2010}, result in ``greener" metallurgy.

The application of three-layer processes was also proposed for
electronic scrap reclamation \cite{SingletonSullivan:1973}, removal of
Mg from scrap Al
\cite{TiwariSharma:1984,GesingDasLoutfy:2016,GesingDas:2017}, and
electrorefining of Si
\cite{OlsenRolseth:2010,OlsenRolsethThonstad:2014,OishiKoyamaTanaka:2016}.
Research on the fluid mechanics of current bearing three-layer systems
can therefore potentially be useful beyond LMBs.

\subsection{Thermally regenerative electrochemical systems}

\begin{figure}
  \centering
  \includegraphics[width=0.5\columnwidth]{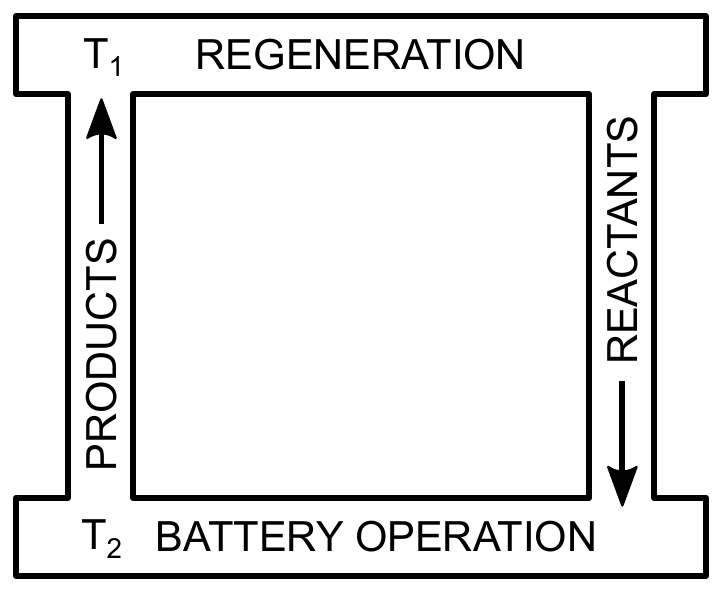}
  \caption{\label{fig:Yeager_1957}Closed cycle battery system suggested by Yeager in 1957, adapted from Roberts \cite{Roberts:1958}.}
\end{figure}

After three-layer liquid metal systems were put to use for Al refining, a few decades passed before they were used to generate electricity. In the meantime, related technologies were developed, including ``closed cycle battery systems'' (Yeager in
\cite{Roberts:1958}), ``thermally regenerative fuel cells'' or
``(thermally) regenerative electrochemical systems (TRES)'' as they
were later subsumed by Liebhafsky \cite{Liebhafsky:1967}, McCully et
al. \cite{MccullyRymarzNicholson:1967}, and Chum and Oster\-young
\cite{ChumOsteryoung:1980,ChumOsteryoung:1981}. TRES combine an
electricity delivering cell with a regeneration unit as sketched in
Fig.~\ref{fig:Yeager_1957}: reactants are combined at the low cell temperature T\textsubscript{2}, and then the product is thermally decomposed at the higher regenerator temperature T\textsubscript{1}. Thermal regeneration implies that the
whole system efficiency is Carnot limited
\cite{Yeager:1958a,Liebhafsky:1959}.

A variety of such systems were investigated in the US during the
period of
1958-1968 \cite{ChumOsteryoung:1980}. 
Later, Chum and Osteryoung classified the published material on this
topic according to system type and thoroughly reviewed it in
retrospect \cite{ChumOsteryoung:1980,ChumOsteryoung:1981}.  LiH based
cells were building blocks of what were probably the first (1958,
\cite{ChumOsteryoung:1980}) experimentally realized thermally
regenerative high-temperature systems \cite{ShearerWerner:1958,CiarlarielloMcdonoughShearer:1961,LawroskiVogelMunnecke:1961}, which continue to be of interest
today
\cite{RoySalamahMaldonadoNarkiewicz:1993,Wietelmann:2014}. Almost at
the same time a patent was filed in 1960 by Agruss
\cite{Agruss:1966},  bimetallic cells were suggested for the electricity
delivering part of TRES. Henderson et al.
\cite{HendersonAgrussCaple:1961} concluded their survey of some 900
inorganic compounds for use in
thermally regenerative fuel cell systems with the recommendation to
concentrate on minimizing electrochemical losses, i.e., polarization
and resistance losses, in order to increase overall efficiency. 
Although unmentioned in \cite{HendersonAgrussCaple:1961}, bimetallic cells with liquid metal electrodes and fused salt
electrolytes were deemed most suitable to fulfill those requirements \cite{AgrussKaras:1962}.
Governmental sponsored research on bimetallic
cells followed soon after at Argonne National Laboratory (1961,
\cite{LawroskiVogelMunnecke:1962}) and at General Motors (1962
\cite{AgrussKaras:1962,Austin:1967}). Research was initially focused
on the application of bimetallic cell based TRES on space power
applications \cite{AgrussKarasDecker:1962}, namely systems using
nuclear reactors as heat sources. Several
studies explored the parameters of concrete designs developed
in frame of the ``Systems Nuclear Auxiliary Power Program'' (SNAP),
SNAP-2 \cite{AgrussKarasDecker:1962,Agruss:1963a,AgrussKaras:1967,Kerr:1967,GroceOldenkamp:1967} and SNAP-8 \cite{ChumOsteryoung:1981}.

Hydrodynamics naturally plays a vital role in the operation of TRES
due to the necessity to transport products and reactants between the electricity producing and the thermal
regeneration parts of the system. However, hydrodynamics of the
transport between the cell and the regenerator is mainly concerned
with the task of pumping \cite{GroceOldenkamp:1967,AgrussKaras:1967}
and the subtleties of keeping a liquid metal flow through~---  while preventing electrical contact between~--- different cells
\cite{AgrussKarasDecker:1962}. Velocities typical for TRES are much
lower then those found in conventional heat engines
\cite{CrouthamelRecht:1967} and could even be achieved using natural
circulation driven by heat
\cite{CairnsCrouthamelFischerEtAl:1967}. Agruss et
al.~\cite{AgrussKarasDecker:1962} emphasized that solutal convection
should be taken into account when designing TRES cells and
flow control is vital to obtain good long-term
performance. Publications covering detailed investigations of cell
specific fluid mechanics are unknown to the present authors, but the
LMB pioneers were obviously aware of its importance as can be seen by
a variety of pertinent notes. 

\begin{figure}
  \centering
  \includegraphics[width=0.95\columnwidth]{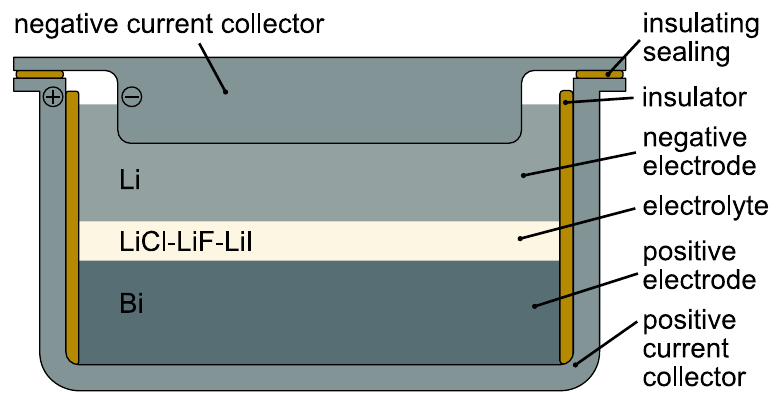}
  \caption{\label{fig:ddc_sketch}Sketch of a differential density cell.}
\end{figure}

Cell construction determines to a large extent the influence
hydrodynamics can have on cell operation. No mechanical obstructions
exist in the ``differential density cell''
\cite{AgrussKarasDecker:1962,Agruss:1963a,AgrussKaras:1967,ChumOsteryoung:1981}
sketched in Fig.~\ref{fig:ddc_sketch}. This is most likely the purest
embodiment of an LMB: inside the cell there are only the three fluid
layers that are floating on top of each other according to
density. Early on, the vital role of stable density stratification was
clearly identified
\cite{WeaverSmithWillmann:1962,ChumOsteryoung:1981,AgrussKaras:1967}.
The interfaces of differential density cells using a K$|$KOH-KBr-KI$|$Hg system
were stable enough to allow for a mild Hg feed of a few milliliters
per minute \cite{AgrussKarasDecker:1962}. Cell performance depended on
the flow distribution, the volume flux, and, vicariously, on the
temperature of the incoming Hg.
Cairns et al.~\cite{CairnsCrouthamelFischerEtAl:1967}
presented a conceptual design for a battery of three Na$|$NaF-NaCl-NaI$|$Bi differential density cells, stressed that the density differences are large enough to clearly separate the phases, and mentioned in the same breath that ``hydrodynamic stability of the liquid streams must be carefully established.''

Restraining one or more of the liquid phases in a porous ceramic
matrix is a straightforward means to guarantee mechanical stability of the interfaces
\cite{Swinkels:1971}. A direct mechanical separation of anodic and
cathodic compartment is a necessity for space applications that could
not rely on gravity to keep the layers apart. Besides the solid
matrix, another means to immobilize electrolytes was to intermix them
with ceramic powders, so called ``fillers'', that resulted in paste
electrolytes. Since both matrix and powders had to be
electric insulators, an overall conductivity reduction by a factor of
about two to four
\cite{ShimotakeRogersCairns:1969,CairnsShimotake:1969} resulted even
for the better paste electrolytes. Obviously, using
mechanically separated electrode compartments is a
prerequisite for any mobile application of LMBs. Equally, for cells
used as components in complete TRES, the constant flow to and from the
regenerator and through the cell necessitates in almost all cases a
mechanical division of positive and negative electrodes. Examples include the
flow through cell with sandwich matrix by Agruss and Karas
\cite{AgrussKaras:1967}, the earlier single cup cell of ``flowing
type'' using an electrolyte impregnated alumina thimble
\cite{Agruss:1963a}, and the paste electrolyte cells developed at
Argonne National Laboratory \cite{VogelProudRoyal:1968,ShimotakeRogersCairns:1969}.

\begin{figure}
  \centering
  \includegraphics[width=0.95\columnwidth]{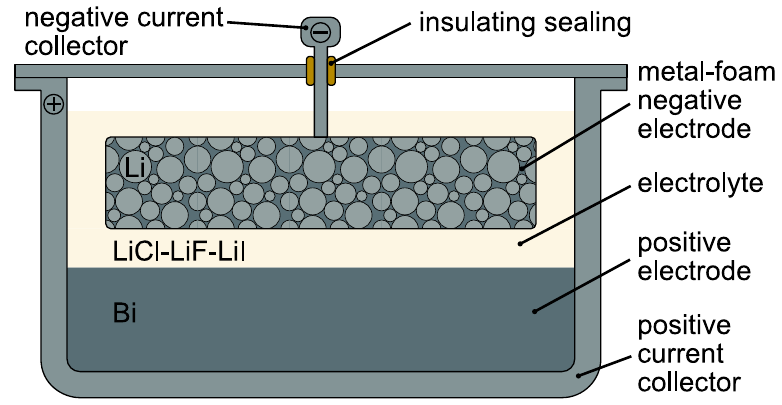}
  \caption{\label{fig:fac_sketch}Sketch of a liquid metal cell
    featuring a retainer (metal foam) to contain the negative electrode.}
\end{figure}

A different purpose was pursued by encasing the negative electrode material into a
retainer \cite{LawroskiVogelLevensonMunnecke:1963} 
made from stainless steel fibers
\cite{CairnsCrouthamelFischerEtAl:1967}, 
felt metal \cite{VogelBurrisTevebaughWebsterProudRoyal:1971}, or later, foam
\cite{WangJiangChungOuchiBurkeBoysenKimMueckeSadoway:2014,
  Ning:2015,SpatoccoOuchiLambotteBurkeSadoway:2015, Ouchi:2016} as
sketched in Fig.~\ref{fig:fac_sketch}. Those retainers allow 
electrical insulation of the negative electrode from the rest of the cell without
resorting to ceramics and restrict fluid mechanics to
that in a porous body. The probably simplest retainer used was an
Armco iron ring \cite{LawroskiVogelMunnecke:1962,CairnsCrouthamelFischerEtAl:1967}
that encased the alkaline metal, a configuration more akin to the
differential density cell than to a porous body. Arrangements similar
to the iron ring are sometimes used as well in molten salt electrolysis cells
\cite{Grube:1930,Drossbach:1952,Gossrau:1955} to keep a patch of molten metal
floating on top of fused salt while preventing contact with
the rest of the cell. In
the case of poorly conducting materials (especially Te and Se),
the positive electrode had to be equipped with additional electronically
conducting components to improve current collection
\cite{VogelProudRoyal:1968,Swinkels:1971}.


With view on the low overall efficiencies of TRES due to Carnot cycle limitations
as well as problems of pumping, plumbing and separation, research on
thermally regenerative systems ceased after 1968
\cite{ChumOsteryoung:1981} and later LMB work at Argonne concentrated on
Li-based systems with chalcogen positive electrodes, namely Se and Te. The high strength of the
bonds in those systems makes them unsuitable for thermal regeneration
\cite{Swinkels:1971}. However, in their review, Chum and Oster\-young
\cite{ChumOsteryoung:1981} deemed it worthwhile to
reinvestigate TRES based on alloy cells once a solar-derived, high
temperature source was identified. Just recently, a Na$||$S based
approaches for solar electricity generation using thermal regeneration
was suggested \cite{WengerEpsteinKribus:2017}.

\subsection{LMBs for stationary storage}

Using bimetallic cells as secondary elements for off-peak electricity
storage was already a topic in the 1960s
\cite{CairnsCrouthamelFischerEtAl:1967,SteunenbergBurries:2000}.  The
very powerful Li$||$Se and Li$||$Te cells \cite{CairnsShimotake:1969} mentioned above are, however, unsuitable for wide scale use
because of the scarcity of the both chalcogens 
\nocite{Bockris:1972} \cite{Swinkels:1971,
  HietbrinkMcbreeSelisTricklebankWitherspoon:1972}. In the late 1960s
and early 1970s, research at Argonne moved on to Li$||$S
\cite{VogelLevensonProudRoyal:1968, Swinkels:1971,HietbrinkMcbreeSelisTricklebankWitherspoon:1972, KyleCairnsWebster:1973}, thus leaving the area of
bimetallic cells.

LMB activities were reinvigorated at MIT in the first years of the
21\textsuperscript{st} century by Donald Sadoway. The initial design conceived in the
fall of 2005 by Sadoway and Ceder and presented by Bradwell
\cite{Bradwell:2006} was one that
combined Mg and Sb with a Mg$_3$Sb$_2$ containing electrolyte,
decomposing the Mg$_3$Sb$_2$ on charge and forming it on
discharge. This new cell design was later termed ``Type A'' or
``ambipolar electrolysis LMB'' \cite{Bradwell:2011} and not followed
up later. Instead, bimetallic alloying cells `(``Type B'') were
investigated using different material combinations whose selection was
not hampered by the need of thermal regeneration. The Mg$||$Sb system
was among the first alloying systems studied at MIT. With
MgCl\textsubscript{2}-NaCl-KCl (50:30:20 mol\%) it used a standard
electrolyte for Mg electrolysis \cite{Ray:2006}. Thermodynamic data
for the system are available from \cite{RaoPatil:1971} and cell
performance data were later reported also by Leung et
al.~\cite{LeungHeckAmietszajewEtAl:2015}.

Right from the start research at MIT focused on the deployment of LMBs
for large-scale energy storage \cite{BradwellKimSirkSadoway:2012}
concentrating on different practical and economical aspects of
utilizing abundant and cheap materials \cite{KimBoysenOuchiEtAl:2013, SpatoccoSadoway:2015}.
Initially, large cells with volumes of a few cubic meters
\cite{SadowayCederBradwell:2012} and cross-sections of ($>$2\,m
$\times$ 2\,m) \cite{Bradwell:2006} were envisioned. Those cells would
have a favorable volume to surface ratio translating into a small
amount of construction material per active material and potentially
decreasing total costs. In addition, in large cells Joule heating in the electrolyte could be sufficient to keep the components molten \cite{BradwellKimSirkSadoway:2012}.

The differential density cells employed for the initial MIT
investigations gave later way to cells that used 
metal foam immersed in the electrolyte to contain the negative electrode 
\cite{WangJiangChungOuchiBurkeBoysenKimMueckeSadoway:2014,Ning:2015,SpatoccoOuchiLambotteBurkeSadoway:2015,Ouchi:2016,OuchiSadoway:2017}.

Successful scaling on the cell level from 1\,cm diameter to 15\,cm
diameter was demonstrated by Ning et al.~\cite{Ning:2015}. Commercial
cells produced by the MIT spin-off Ambri have square cross-sections of 10\,cm and
20\,cm edge length \cite{Sadoway:2016}. Thus state-of-the-art cells
are moderately sized, but the quest for large scale cells is
ongoing. Recently, Bojarevi{\v c}s et
al.~\cite{BojarevicsTucsPericleous:2016,BojarevicsTucs:2017} suggested
to retrofit old aluminium electrolysis potlines into large scale LMB
installations ending up with cells of 8\,m by 3.6\,m cross-section and
about 0.5\,m total liquid height.

It should be stressed that LMBs form a whole category of battery
systems comprising a variety of material combinations. Consequently,
depending on the active materials and the electrolyte selected,
different flow situations may arise even under identical geometrical
settings.

\section{Thermal Convection and Magnetoconvection}
\label{sec:convection}

Because almost every known fluid expands when heated, spatial variations (gradients) in temperature cause gradients in density. In the presence of gravity, if those gradients are large enough, denser fluid sinks and lighter fluid floats, causing thermal convection. Usually large-scale convection rolls characterize the flow shape. Being ubiquitous and fundamental in engineering and natural systems, convection has been studied extensively, and many reviews of thermal convection are available~\cite{Chilla:2012,Lohse:2010,Ahlers:2009,Bodenschatz:2000}. Here we will give a brief introduction to convection, then focus on the particular characteristics of convection in liquid metal batteries. Joule heating drives convection in some parts of a liquid metal battery, but inhibits it in others. Broad, thin layers are common, and convection in liquid metal batteries differs from aqueous fluids because metals are excellent thermal conductors (that is, they have low Prandtl number). Convection driven by buoyancy also competes with Marangoni flow driven by surface tension, discussed in \S\ref{sec:Marangoni}. We close this section with a discussion of magnetoconvection, in which the presence of magnetic fields alters convective flow. 

\subsection{Introduction to thermal convection}

Thermal convection is driven by gravity, temperature gradients, and thermal expansion, but hindered by viscosity and thermal diffusion. Convection also occurs more readily in thicker fluid layers. Combining these physical parameters produces the dimensionless Rayleigh number
\[
Ra = \frac{g \alpha_T \Delta T L^3}{\nu \kappa},
\]
where $g$ is the acceleration due to gravity, $\alpha_T$ is the
coefficient of volumetric expansion, $\Delta T$ is the characteristic
temperature difference, $L$ is the vertical thickness, $\nu$ is the
kinematic viscosity, and $\kappa$ is the thermal diffusivity. The
Rayleigh number can be understood as a dimensionless temperature
difference and a control parameter; for a given fluid and vessel
shape, convection typically begins at a critical value
$Ra>Ra_\mathrm{crit}>0$, and subsequent instabilities that change the
flow are also typically governed by $Ra$. 

If the Rayleigh number is understood as a control parameter, then the results of changing $Ra$ can also be expressed in terms of dimensionless quantities. The Reynolds number
\begin{equation}
    \label{eq:Rayleigh}
Re = \frac{U L}{\nu},
\end{equation}
where $U$ is a characteristic flow velocity, can be understood as a dimensionless flow speed. The Nusselt number
\[
Nu = -\frac{QL}{\kappa\Delta T},
\]
where $Q$ is the total heat flux through the fluid, can be understood as a dimensionless heat flux. 

The canonical and best-studied context in which convection occurs is the Rayleigh-B\'enard case, in which a fluid is contained between upper and lower rigid, no-slip boundaries, with the lower boundary heated and the upper boundary cooled. Usually both boundaries are held at steady, uniform temperatures, or subjected to steady, uniform heat flux. Convection also occurs in many other geometries, for example lateral heating. Heating the fluid from above, however, produces a stably-stratified situation in which flow is hindered. 

\subsection{Introduction to compositional convection}

Temperature is not the only parameter that affects fluid density. Chemical reactions, for example, can also change the local density such that buoyancy drives flow. That process is known as compositional convection, and the corresponding control parameter is the compositional Rayleigh number
\[
Ra_X = \frac{g \alpha_X \Delta X L^3}{\nu D},
\]
where $\alpha_X$ is the coefficient of volumetric expansion with concentration changes, $\Delta X$ is the characteristic concentration difference, and $D$ is the material diffusivity. (Compositional convection is one mechanism by which reaction drives flow; entropic heating, discussed above, is another.) For liquid metal batteries, the electrode materials have densities that differ by more than an order of magnitude (see Table~\ref{tab:electrodeProperties}), and $\Delta X \sim 30$~mol\%, so we expect compositional convection to cause substantial flow. For comparison, we can consider thermal convection in bismuth at 475\,$\mbox{}^\circ$C, for which the coefficient of thermal expansion is $\alpha_T=1.24 \times 10^{-4}$/K~\cite{NEA:2015}. Making the order-of-magnitude estimate $\Delta T \sim 1$~K, it becomes clear that $\alpha_X \Delta X \gg \alpha_T \Delta T$. For the Na$||$Bi system at an operating temperature of 475\,$\mbox{}^\circ$C, the compositional Rayleigh number exceeds the thermal one by six orders of magnitude. Thus compositional convection is likely much stronger than thermal convection. Compositional convection is unlikely during discharge because the less-dense negative electrode material (e.g., Li) is added to the top of the more-dense positive electrode, producing a stable density stratification. During charge, however, less-dense material is removed from the top of the positive electrode, leaving the remaining material more dense and likely to drive compositional convection by sinking. 

\begin{table*}[t]
    \caption{\label{tab:electrodeProperties} Properties of common
      electrode materials. Values for metals at respective melting
      temperature taken from \cite{IidaGuthrie:2015,IidaGuthrie:2015a}
      except Li conductivity from \cite{Davidson:1968}, Pb-Bi eutectic
      data from Sobolev \cite{Sobolev:2011}, and Pb data
      from~\cite{NEA:2015}.}
    \center
    \begin{tabular}{llllllll}
        \hline
        \hline
        \textbf{Material} & $\nu/10^{-6}$ (m\textsuperscript{2}/s) & $\kappa/10^{-5}$ (m\textsuperscript{2}/s) & $\rho$ (kg/m\textsuperscript{3}) & $\alpha_T/10^{-4}$ (K\textsuperscript{-1}) & $\sigma_E/10^6$ (S/m) & $Pr$ & $Pm/10^{-6}$ \\
        \hline
        \multicolumn{8}{l}{\emph{Negative electrode}} \\
        Li & 1.162 & 2.050 & 518 & 1.9 & 3.994 & 0.0567 & 5.8315 \\
        Mg & 0.7862 & 3.4751 & 1590 & 1.6 & 3.634 & 0.0226 & 3.5907 \\
        Na & 0.7497 & 6.9824 & 927 & 2.54 & 10.42 & 0.0107 & 9.8195 \\
        \hline
      \multicolumn{8}{l}{\emph{Positive electrode}} \\
        Bi & 0.1582 & 1.1658 & 10050 & 1.17 & 0.768 & 0.0136 & 0.1527 \\
        Pb & 0.253 & 1.000 & 10673 & 1.199 & 1.050 & 0.0253 & 0.334 \\ 
        Sb & 0.2221 & 1.3047 & 6483 & 1.3 & 0.890 & 0.0170 & 0.2485 \\
        Zn & 0.5323 & 1.5688 & 6575 & 1.5 & 2.67 & 0.0339 & 1.7860 \\
        eutectic Pb-Sb & $\nu$ & $\kappa$ & $\rho$ & $\alpha_T$ & $\sigma_E$ & $Pr$ & $Pm$ \\
        eutectic Pb-Bi & 0.3114 & 0.5982 & 10550 & 1.22 & 0.909 & 0.052 & 0.3557\\
        \hline
        \hline
    \end{tabular}
\end{table*}

\subsection{Convection in liquid metal batteries}

\begin{figure}
  \centering
  \includegraphics[width=\columnwidth]{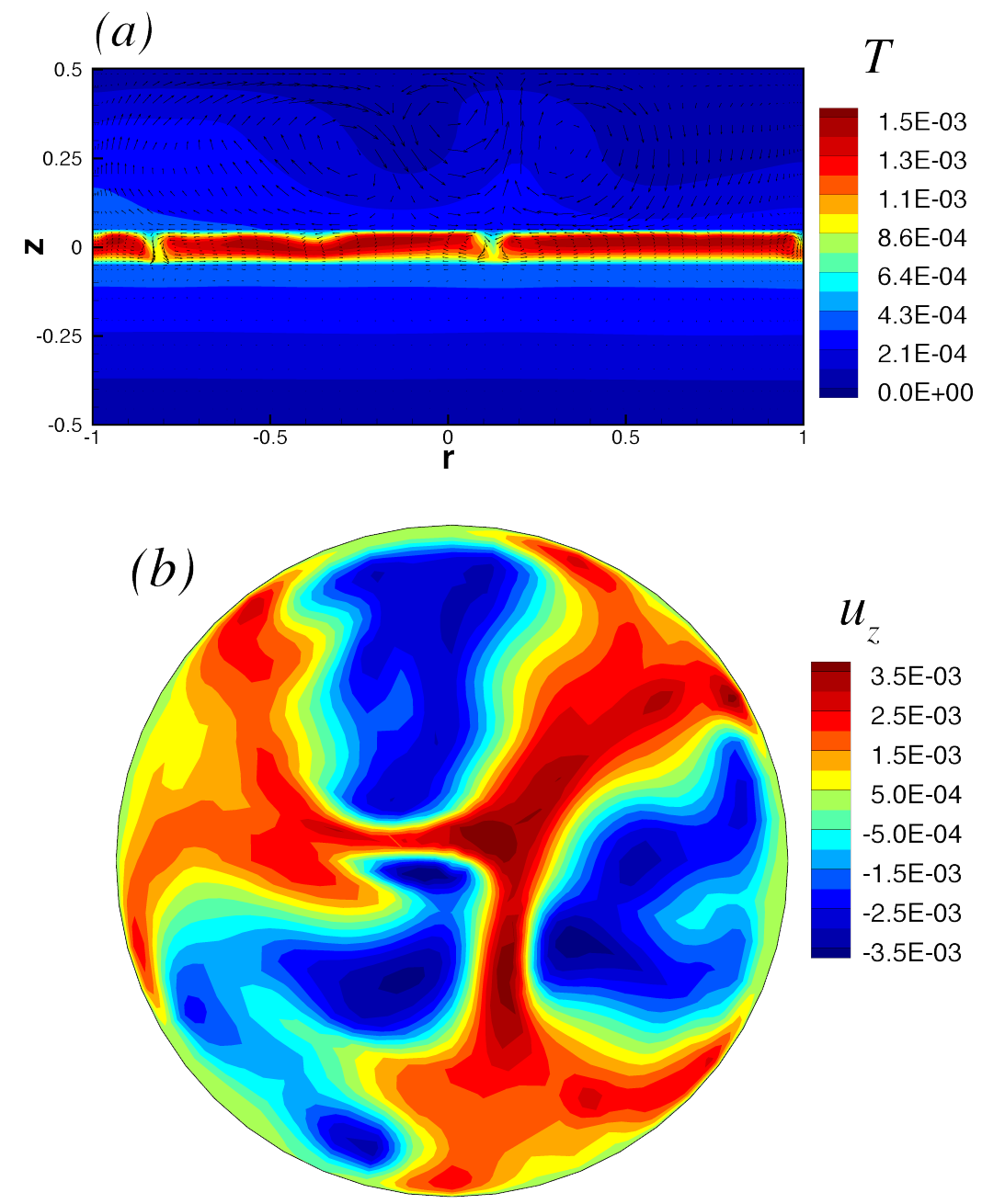}
  \caption{\label{fig:ShenZikanov2016fig8ad}Thermal convection in a three-layer liquid metal battery. A vertical cross-section through the center of the battery (a) shows that the temperature is much higher in the electrolyte than in either electrode. A horizontal cross-section above the electrolyte (b) shows vigorous flow. (Here $u_z$ is the vertical velocity component.) Adapted from~\cite{ShenZikanov:2016}, with permission.}
  \end{figure}

\begin{figure}
  \centering
  \includegraphics[width=0.8\columnwidth]{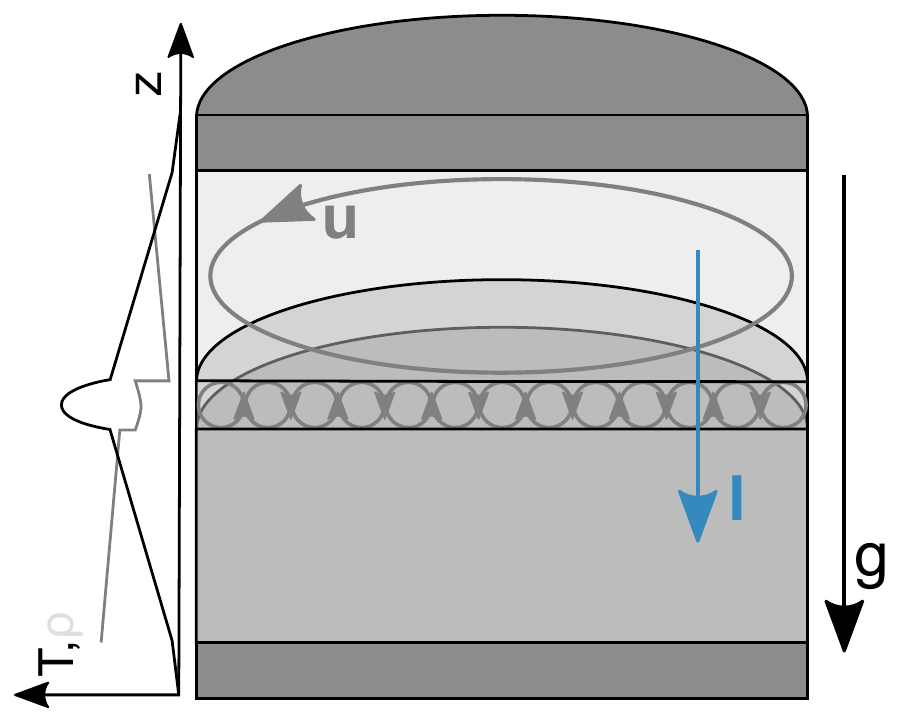}
  \caption{\label{fig:lmb_convection}Sketch of a liquid metal cell with thermal convection}
  \end{figure}

Liquid metal batteries as sketched in
  Fig.~\ref{fig:lmb_convection} are a more complicated and
interesting case than a single layer
  system. Commercially viable liquid metal
battery chemistries involve materials that are solid at room
temperature; to operate, they must be heated to
475\,$\mbox{}^\circ$C~\cite{WangJiangChungOuchiBurkeBoysenKimMueckeSadoway:2014}. External
heaters produce thermal convection in almost any arrangement,
especially the most efficient one in which heaters are installed below
the batteries, producing the Rayleigh-B\'enard case. During operation,
however, external heaters are often unnecessary because the electrical
resistance of the battery components converts electrical energy to
heat in a process known as Joule heating or ohmic heating. If the battery current is large enough and the environmental heat loss is small enough, batteries can maintain temperature without additional heating~\cite{Barriga:2013}. (In fact, cooling may sometimes be necessary.) In this case, the primary heat source lies not below the battery, but within it. As Table~\ref{tab:electrolyteProperties} shows, molten salts have electrical conductivity typically four orders of magnitude smaller than liquid metals, so that essentially all of the Joule heating occurs in the electrolyte layer, as shown in Fig.~\ref{fig:ShenZikanov2016fig8ad}. The positive electrode, located below the electrolyte, is then heated from above and becomes stably stratified; its thermal profile actually hinders flow. Some flow may be induced by the horizontal motion of the bottom of the electrolyte layer, which slides against the top of the positive electrode and applies viscous shear stresses, but simulations of Boussinesq flow show the effect to be weak~\cite{ShenZikanov:2016,ZikanovShen:2016}. The electrolyte itself, which experiences substantial bulk heating during battery charge and discharge, is subject to thermal convection, especially in its upper half~\cite{ZikanovShen:2016,KoellnerBoeckSchumacher:2017}. One simulation of an internally-heated electrolyte layer showed it to be characterized by small, round, descending plumes~\cite{XiangZikanov:2017}. Experiments have raised concern that thermal convection could bring intermetallic materials from the electrolyte to contaminate the negative electrode~\cite{Foster:1967}. Convection due to internal heating has also been studied in detail in other contexts~\cite{Goluskin:2015}.

\begin{table*}[t]
    \caption{\label{tab:electrolyteProperties} Properties of common
      electrolyte materials, from Janz \textit{et
        al.}~\cite{Janz:1968,Janz:1979} Todreas et
      al.~\cite{TodreasHejzlarFongNikiforovaPetroskiShwagerausWhitman:2008},
    Kim et al.~\cite{KimBoysenNewhouseEtAl:2013}, and Masset et
    al.~\cite{MassetHenryPoinsoPoignet:2006,MassetSchoeffertPoinsoPoignet:2005a}.}
    \center
    \begin{tabular}{llll}
        \hline
        \hline
        \textbf{Material} & $\nu/10^{-6}$ (m$^2$/s) & $\rho$ (kg/m$^3$) & $\sigma_E$ (S/m) \\
        \hline
      LiF & 1.228 & 1799 & 860 \\ 
        LiCl & 1.067 & 1490 & 586 \\ 
        LiI & 0.702 & 3.0928 & 396.68 \\ 
        NaCl & 0.892 & 1547 & 363 \\ 
        CaCl$_2$ & 1.607 & 2078 & 205.9 \\ 
      BaCl$_2$ & 1.460 & 3.150 & 216.4 \\ 
      NaOH & 2.14 & 1767 & 244 \\ 
        NaI & 0.532 & 2725.8 & 229.2 \\ 
      ZnCl$_2$ & 1150 & 2514 & 0.268 \\ 
      LiCl-KCl & 1.560 & 1563 & 157.2 \\[-0.7em] 
      (58.5-41.5) mol\% & & & \\
      NaCl-KCl-MgCl$_2$ & 0.688 & 1715 & 80 \\[-0.7em] 
      (30-20-50) mol\% & & & \\
      LiCl-LiF-LiI & - & 2690 & 288 \\[-0.7em] 
      (29.1-11.7-59.2) mol\% & & &\\
        \hline
        \hline
    \end{tabular}
\end{table*}

The negative electrode, located above the electrolyte, is heated from
below and is subject to thermal convection. In a negative electrode
composed of bulk liquid metal, we would expect both unstable thermal
stratification and viscous coupling to the adjacent electrolyte to
drive flow. Simulations show that in parameter regimes typical of
liquid metal batteries, it is viscous coupling that dominates; flow
due to heat flux is negligible~\cite{ShenZikanov:2016}. Therefore, in
the case of a thick electrolyte layer, mixing in the electrolyte is
stronger than mixing in the negative electrode above; in the case of a
thin electrolyte layer, the roles are reversed~\cite{ShenZikanov:2016}. However, negative electrodes may also be held in the pores of a rigid metal foam by capillary forces, which prevents the negative electrode from contacting the battery sidewall~\cite{WangJiangChungOuchiBurkeBoysenKimMueckeSadoway:2014}. The foam also substantially hinders flow within the negative electrode. Essentially the characteristic length scale becomes the pore size of the foam, which is much smaller than the thickness of the negative electrode. Since the Rayleigh number is proportional to the cube of the characteristic length scale (Eq.~\ref{eq:Rayleigh}), convection is drastically weakened, if not prevented altogether. The physics of convection in porous media~\cite{Shattuck:1997,Howle:1997} might apply in this case. 

If a liquid metal battery is operated with current density that is uniform across is horizontal cross-section, we expect uniform Joule heating and therefore temperatures that vary primarily in the vertical direction (aside from thermal edge effects). However, the negative current collector must not make electrical contact with the vessel sidewall, which is part of the positive current collector. For that reason, the metal foam negative current collector that contains the negative electrode is typically designed to be smaller than the battery cross-section, concentrating electrical current near the center and reducing it near the sidewall. The fact that current can exit the positive electrodes through the sidewalls as well as the bottom wall allows further deviation from uniform, axial current. Nonuniform current density causes Joule heating that is also nonuniform~--- in fact, it varies more sharply, since the rate of heating is proportional to the square of the current density. This gradient provides another source of convection-driven flow. Putting more current density near the central axis of the battery creates more heat there and causes flows that rise along the central axis. Interestingly electro-vortex flow (considered in detail in \S\ref{sec:electro-vortex}) tends to cause the opposite motion: descent along the central axis. Simulations have shown that negative current collector geometry and conductivity substantially affect flow in liquid metal batteries~\cite{WeberGalindoPriedeEtAl:2015}. Other geometric details can also create temperature gradients and drive convection. For example, sharp edges on a current collector concentrate current and cause intense local heating. The resulting local convection rolls are small but can nonetheless alter the global topology of flow and mixing. Also, if solid intermetallic alloys form, they affect the boundary conditions that drive thermal convection. Intermetallics are typically less dense than the surrounding melt, so they float to the interface between the positive electrode and electrolyte. Intermetallics typically have lower thermal and electrical conductivity than the melt, so where they gather, both heating and heat flux are inhibited, changing convection in non-trivial ways. 

\subsection{Metals and salts: Convection at low Prandtl number}

In addition to the Rayleigh number, a second dimensionless parameter specifies the state of a convecting system, the Prandtl number
\[
Pr = \frac{\nu}{\kappa}.
\]
A ratio of momentum diffusivity (kinematic viscosity) to thermal diffusivity, the Prandtl number is a material property that can be understood as a comparison of the rates at which thermal motions spread momentum and heat. Table~\ref{tab:electrodeProperties} lists the Prandtl number of a few relevant fluids. Air and water are very often the fluids of choice for convection studies, since so many industrial and natural systems involve them. But air and water have Prandtl numbers that differ from liquid metals and molten salts by orders of magnitude: $Pr=7$ for water and $Pr=0.7$ for air. We therefore expect thermal convection in liquid metals and molten salts to differ substantially from convection in water or air. 

The Prandtl number plays a leading role in the well-known scaling
theory characterizing turbulent convection, developed by Grossmann and
Lohse~\cite{Grossmann:2000a}. In fact, the scaling theory expresses
the outputs $Re$ and $Nu$ in terms of the inputs $Ra$ and $Pr$. To
begin, every possible Rayleigh-B\'enard experiment is categorized
according to the role of boundary layers in transporting momentum and
heat. Boundary layers occur near walls, and transport through them
proceeds (to a good approximation) by diffusion alone. On the other
hand, in the bulk region far from walls, transport proceeds primarily
by the fast and disordered motions typical in turbulent flow. Any
particular Rayleigh-B\'enard experiment can be assigned to one of
eight regimes, depending on three questions: Is momentum transport
slower through the boundary layer or the bulk? Is heat transport
slower through the boundary layer or the bulk? And, which boundary
layer~--- viscous or thermal~--- is thicker and therefore dominant?
Answering those three questions makes it possible to estimate the
exponents that characterize the dependence of $Re$ and $Nu$ on $Ra$
and $Pr$. According to the theory~\cite{Grossmann:2000a}, the Nusselt
number can depend on the Prandtl number as weakly as $Nu \propto
Pr^{-1/12}$ or as strongly as $Nu \propto Pr^{1/2}$, and the Reynolds
number can depend on the Prandtl number as weakly as $Re \propto
Pr^{-1/2}$ or as strongly as $Re \propto Pr^{-6/7}$. Again, convection
in liquid metals and molten salts differs starkly from convection in
water or air: changing $Pr$ by orders of magnitude causes $Re$ and
$Nu$ to change by orders of magnitude as well. Experiments studying
convection in sodium ($Pr=0.0107$) have confirmed that the heat flux ($Nu$) for a given temperature difference ($Ra$) is indeed smaller than for fluids with larger $Pr$~\cite{Horanyi:1999}. Experiments have also shown that at low $Pr$, more of the flow's kinetic energy is concentrated in large-scale structures, especially large convection rolls. In a thin convecting layer with a cylindrical sidewall resembling the positive electrode of a liquid metal battery, slowly fluctuating concentric ring-shaped rolls often dominate~\cite{Horanyi:1999}. Those rolls may interact via flywheel effects~\cite{Jones:1976}. 

\subsection{Magnetoconvection}

Convection in liquid metal batteries proceeds in the presence of~--- and can be substantially altered by~--- electric currents and magnetic fields. Introductions and overviews of the topic of magnetoconvection have been provided in texts dedicated to the subject~\cite{Weiss:2014} as well as texts on the more general topic of magnetohydrodynamics~\cite{Davidson:2001}. The strength of the magnetic field can be represented in dimensionless form using the Hartmann number
\begin{equation}
\label{eq:Hartmann}
Ha = B L \sqrt{\frac{\sigma_E}{\rho \nu}},
\end{equation}
which is the ratio of electromagnetic force to viscous force. Here $B$ is the characteristic magnetic field strength, $\sigma_E$ is the electrical conductivity, and $\rho$ is the density. (Magnetic field strength is also sometimes expressed using the Chandrasekhar number, which is the square of the Hartmann number.) When $Ha \gg 1$, magnetic fields tend to strongly alter convection, though the particular effects depend on geometry. 

When an electrically conducting fluid flows in the presence of a magnetic field, electrical currents are induced, and those currents in turn produce magnetic fields. According to Lenz's law, the direction of any induced current is such that it opposes change to the magnetic field. Accordingly, conductive fluids flow most easily in directions perpendicular to the local magnetic field. The simplest such flows form paths that circulate around magnetic fields; in the presence of magnetic fields, convection rolls tend to align with magnetic field lines. That phenomenon is analogous to the tendency of charged particles in plasmas to orbit magnetic field lines. Other motions, such as helical paths, are also possible. 

If convection occurs in the presence of a vertical magnetic field,
alignment is impossible, since convection rolls are necessarily
horizontal. Accordingly, vertical magnetic fields tend to damp
convection~\cite{Chandrasekhar:1954,Busse:1982,Burr:2001}. The
critical Rayleigh number at which convection begins scales as
$Ra_\mathrm{crit} \propto Ha^2$~\cite{Chandrasekhar:1954}, as has been
verified experimentally~\cite{Aurnou:2001}. The Rayleigh number of oscillatory instability of convection rolls is also increased by the presence of a vertical magnetic field~\cite{Busse:1982}. Common sources of vertical magnetic fields in liquid metal batteries include the Earth's field (though it is relatively weak) and fields produced by wires carrying current to and from the battery. 

Just as the Grossmann and Lohse scaling theory~\cite{Grossmann:2000a} considers the dependence of $Re$ and $Nu$ on the inputs $Ra$ and $Pr$ in convection without magnetic fields, a recent scaling theory by Schumacher and colleagues~\cite{Zurner:2016} considers the dependence of $Re$ and $Nu$ on the inputs $Ra$, $Pr$~--- and also $Ha$~--- in the presence of a vertical magnetic field. The reasoning is analogous: the scaling depends on whether transport time is dominated by the boundary layer or the bulk, and which boundary layer is thickest. However, the situation is made more complex by the need to consider magnetic field transport in addition to momentum and temperature transport, and the possibility that the Hartmann (magnetic) boundary layer might be thickest. Altogether, 24 regimes are possible. To reduce the number of free parameters, the authors considered the case in which $Pr \ll 1$ and $Pm \ll 1$, where $Pm = \nu \mu \sigma_E$ is the magnetic Prandtl number. (Here $\mu$ is the magnetic permeability.) That special case applies to materials common in liquid metal batteries, and still spans four regimes of magnetoconvection. Categorization depends on whether the magnetic field is strong ($Ha \gg 1$), and whether the flow is substantially nonlinear ($Ra \gg 1$). Scaling laws are proposed for each regime. The theory's fit parameters remain unconstrained in three of the four regimes because appropriate experimental data are unavailable. Experiments to produce those data would substantially advance the field. 

On the other hand, if a horizontal magnetic field is present, convection rolls are often able to align with it easily. In that case, flow speed ($Re$) and heat flux ($Nu$) increase~\cite{Burr:2002}. Moreover, since magnetic fields of any orientation damp turbulence~\cite{Moffatt:1967}, convection in the presence of horizontal magnetic fields tends to be more ordered, spatially, than convection in the $Ha=0$ case. As $Ra$ increases, waves develop on the horizontal convection rolls~\cite{Busse:1982,Horanyi:1999}.

In liquid metal batteries, internal electrical currents run primarily vertically and induce toroidal horizontal magnetic fields. Poloidal convection rolls are therefore common, since their flow is aligned and circulates around the magnetic field lines. If the sidewall is cylindrical, boundary conditions further encourage poloidal convection rolls. Such rolls have been observed in liquid metal battery experiments, and the characteristic mass transport time decreases as $Ha$ increases~\cite{KelleySadoway:2014,PerezKelley:2015}. Simulations have shown similar results, with the number of convection rolls decreasing as the current increases~\cite{Beltran:2017}. Other simulations, however, have suggested that electromagnetic effects are negligible for liquid metal batteries with radius less than 1.3~m~\cite{ShenZikanov:2016,ZikanovShen:2016}. Further study may refine our understanding. In batteries with a rectangular cross-section, we would expect horizontal convection rolls circulating around cores that are nearly circular near the central axis of the battery, where the magnetic field is strong and the sidewall is remote. Closer to the wall, we would expect rolls circulating around cores that are more nearly rectangular, due to boundary influence. 

\section{Marangoni flow}
\label{sec:Marangoni}

The molecules of a stable fluid are typically attracted more strongly to each other than to other materials. The result is the surface tension (or surface energy) $\sigma$, which can be understood as an energy per unit area (or a force per unit length) of interface between two materials. The surface tensions of liquid metals and molten salts are among the highest of any known materials, so it is natural to expect surface tension to play a role in liquid metal batteries. This section will consider that role. 

\begin{figure}
  \centering
  \includegraphics[width=0.8\columnwidth]{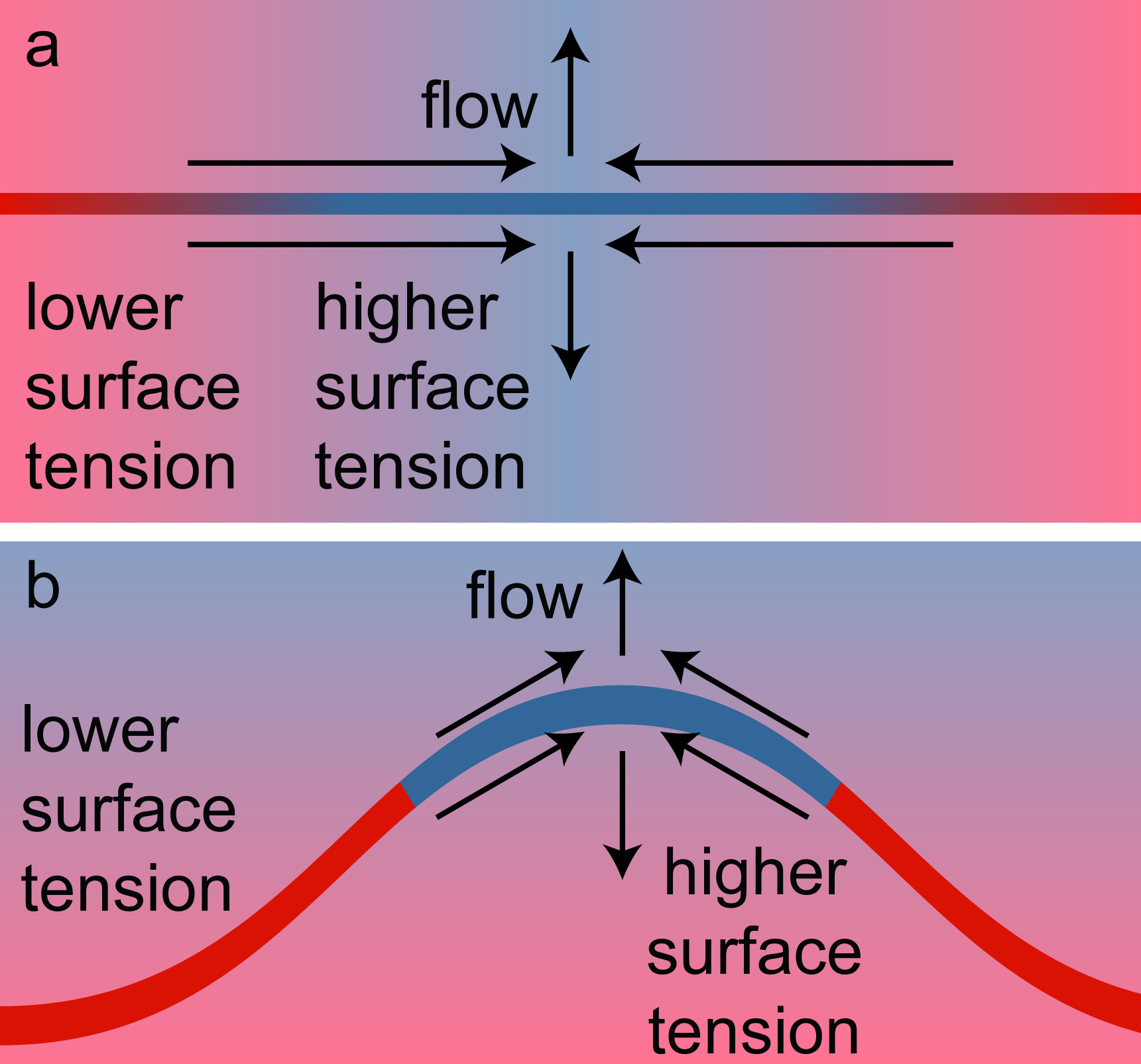}
  \caption{\label{fig:Marangoni}Marangoni flow occurs when surface tension at a fluid interface varies spatially. Variation along the interface (a) always drives flow. Variation across the interface (b) causes an instability that drives flow if the variation is sufficiently large, as quantified by the Marangoni number $Ma$.}
  \end{figure}

If the surface tension varies spatially, regions of higher surface tension pull fluid along the interface from regions of lower surface tension. Viscosity couples that motion to the interior fluid, causing ``Marangoni flow'', sketched in Fig.~\ref{fig:Marangoni}. Surface tension can vary spatially because it depends on temperature, chemical composition, and other quantities. For most fluids, surface tension decreases with temperature: $\partial\sigma/\partial T < 0$, and flow driven by the variation of surface tension with temperature is called ``thermocapillary flow'' and is described in existing reviews~\cite{Davis:1987,Schatz:2001}. Flow driven by the variation of surface tension with composition, called ``solutal Marangoni flow'', has also been considered~\cite{Bratsun:2004,Koellner:2013,Koellner:2015}, especially in the context of thin films~\cite{Jensen:1991,Craster:2009}. In this section we will focus on Marangoni flow phenomena that are relevant to liquid metal batteries, focusing on similarities and differences to Marangoni flows studied in the past. We will estimate which phenomena are likely to arise, drawing insight from one pioneering study that has considered the role of thermocapillary flow in liquid metal batteries~\cite{KoellnerBoeckSchumacher:2017}. 

\subsection{Introduction to thermocapillary Marangoni flow}

Thermocapillary flow is driven by temperature gradients but hindered by viscosity, thermal diffusion, and density (which provides inertia). Thermocapillary flow also occurs more readily in thicker fluid layers. Combining these physical parameters produces the dimensionless Marangoni number
\begin{equation}
\label{eq:Marangoni}
Ma = \frac{\left| \frac{\partial\sigma}{\partial T} \right| L \Delta T}{\rho \nu \kappa}.
\end{equation}
The Marangoni number plays a role analogous to the Rayleigh number in thermal convection. Larger values of $Ma$ make thermocapillary flow more likely and more vigorous. Because temperature gradients drive both thermocapillary flow and thermal convection, the two phenomena often occur simultaneously. We can compare their relative magnitudes via the dynamic Bond number
\begin{equation}
\label{eq:Bond}
Bo = \frac{Ra}{Ma} = \frac{\rho \alpha_T g L^2}{\left| \frac{\partial \sigma}{\partial T} \right|}. 
\end{equation}
Thermal convection dominates when $Bo \gg 1$, whereas thermocapillary flow dominates when $Bo \ll 1$. Because of the $L^2$ factor, thermal convection tends to dominate in thick layers, whereas thermocapillary flow tends to dominate in thin layers. Thermocapillary flow, like thermal convection, is qualitatively different for fluids with small Prandtl number (like liquid metals and molten salts) than for fluids with large Prandtl number. 

\subsection{Surface tension variation across the interface}

Thermocapillary flow phenomena depend on the direction of the thermal gradient with respect to the interface. Surface tension varying \emph{along} the interface always drives flow, as shown in Fig.~\ref{fig:Marangoni}a. We will consider this case in greater detail below. However, if temperature (and thus surface tension) varies \emph{across} the interface, as in Fig.~\ref{fig:Marangoni}b, the situation is more complicated. Thermocapillary flow is possible only if heat flows across the interface in the direction of increasing thermal diffusivity~\cite{KoellnerBoeckSchumacher:2017}. That condition is satisfied at both interfaces between electrolyte and electrode in a liquid metal battery because the molten salt electrolyte has lower thermal diffusivity than the metals, and typically higher temperature as well, because of its low electrical conductivity. 

With the directional condition satisfied, three phenomena are possible~\cite{Davis:1987,Schatz:2001}. First, the fluid can remain stagnant if thermal conduction carries enough heat and the viscosity is large enough to damp flow. Second, short-wavelength thermocapillary flow can arise, in which the surface deformations caused by surface tension are damped primarily by gravity. Third, long-wavelength thermocapillary flow can arise, in which the surface deformations caused by surface tension are damped primarily by diffusion (of both momentum and heat). The relative strength of the two damping mechanisms is quantified by the Galileo number
\begin{equation}
\label{eq:Galileo}
G = \frac{g d^3}{\nu \kappa}. 
\end{equation}
$G \gg 1$ implies that gravity is the primary damping mechanism, such that we expect short-wavelength flow, whereas $G \ll 1$ implies that diffusion is the primary damping mechanism, such that we expect long-wavelength flow~\cite{VanHook:1995}. (Note that some past authors~\cite{Davis:1987,Schatz:2001} have used the term ``Marangoni flow'' for the specific case of thermocapillary flow driven by temperature variation across the interface, not for the much more general case of all flows driven by surface tension, as we use it here.) 

In the $G \gg 1$ case, linear stability theory shows that conductive heat transfer becomes unstable and short-wavelength flow arises when $Ma >80$~\cite{Pearson:1958}, as experimental studies have confirmed~\cite{VanHook:1995}. Typically short-wavelength flow appears as an array of hexagons tiled across the interface. For $G \le 120$, linear stability theory predicts that when $Ma>2G/3$, instead of short-wavelength flow, long-wavelength flow arises~\cite{Smith:1966,VanHook:1995}. The long-wavelength flow has no repeatable or particular shape, instead depending sensitively on boundary conditions. When observed in experiments, the long-wavelength flow always ruptures the layer in which it occurs~\cite{Schatz:2001}, a property particularly alarming for designers of liquid metal batteries. The short-wavelength mode, on the other hand, causes nearly zero surface deformation~\cite{Schatz:2001}. 

We can estimate the relevance of thermocapillary flow and the likelihood of short-wavelength and long-wavelength flow using dimensionless quantities, as long as the necessary material properties are well-characterized. Most difficult to obtain is the rate of change of surface tension with temperature, $\partial \sigma/\partial T$. Its value is well-known for Pb, Bi, and their eutectic alloy~\cite{NEA:2015} because of its importance in nuclear power plants, however. One pioneering study~\cite{KoellnerBoeckSchumacher:2017} simulated thermocapillary flow in a hypothetical three-layer liquid metal battery with a eutectic PbBi positive electrode, a LiCl-KCl electrolyte, and a Li negative electrode. First considering the positive electrode, for a PbBi layer with $L=20$~mm and $\Delta T=0.5$~K (the conditions used in the study), we expect no short-wavelength flow because, according to eq.~\ref{eq:Marangoni}, $Ma = 14 < 80$. Nor do we expect long-wavelength flow because, according to eq.~\ref{eq:Galileo}, $G = 6 \times 10^7 \gg 120$. Now considering a LiCl-KCl electrolyte layer with $L=20$~mm and $\Delta T=6$~K (again matching~\cite{KoellnerBoeckSchumacher:2017}), we come to different conclusions: $Ma = 23,000$ implies vigorous thermocapillary flow, and $Bo = 9$ leads us to expect thermocapillary flow of speed similar to the thermal convection. Because $G = 4.7 \times 10^6$, we expect the short-wavelength mode, not the long-wavelength mode. Finally, we expect minimal flow in the negative electrode if it is contained in a rigid metal foam. All of these predictions should be understood as preliminary since eqs.~\ref{eq:Marangoni}, \ref{eq:Bond}, and \ref{eq:Galileo} consider a layer in which only one surface is subject to surface tension effects, but the electrolyte layer in a liquid metal battery is subject to surface tension effects on both is upper and lower surfaces. 

In fact, though the long-wavelength mode can readily be observed in laboratory experiments with silicone oils~\cite{Koschmieder:1992,VanHook:1995,VanHook:1997}, liquid metals and molten salts typically have much smaller kinematic viscosity and thermal diffusivity, yielding small values of $G$ that make the long-wavelength mode unlikely. Using the $Ma > 2G/3$ criterion and the appropriate material properties, we find that the long-wavelength mode will appear only for thicknesses $60~\mu$m or less in either the PbBi or LiCl-KCl layer. Other considerations require both the electrolyte and the positive electrode to be much thicker, so rupture via the long-wavelength thermocapillary mode is unlikely in a liquid metal battery. 

We would expect, however, that the short-wavelength thermocapillary mode often arises in liquid metal batteries, especially in the electrolyte layer. Though unlikely to rupture the electrolyte, the short-wavelength mode may mix the electrolyte, promoting mass transport. The short-wavelength mode might also couple to other phenomena, for example the interfacial instabilities discussed in \S\ref{sec:interface}.

\subsection{Surface tension variation along the interface}

Surface tension that varies \emph{along} the interface always drives flow, and we can estimate its speed by considering the energy involved. Suppose a thin, rectangular layer of fluid occupies the region $0 \le x \le L_x$, $0 \le y \le L_y$, $0 \le z \le L_z$ in Cartesian coordinates $(x,y,z)$, with $L_z \ll L_x$ and $L_z \ll L_y$. Suppose thermocapillary forces act on the $z=L_z$ surface, and that temperature varies in the $x$ direction, such that surface tension drives flow in the $x$ direction. The work done by thermocapillary forces (per unit volume) scales as
\[
\frac{\Delta T \frac{\partial \sigma}{\partial T} L_x L_y}{L_x L_y L_z}.
\]
If the flow is steady, if pressure variations are negligible, and if inertial and gravitational forces are negligible, then the work done by thermocapillary forces must be dissipated by viscous damping. For an incompressible Newtonian fluid, the viscous damping term (in energy per unit volume) reads
\[
\mu \tau u_j \left( \frac{\partial^2 u_i}{\partial x_i \partial x_j} + \frac{\partial^2 u_j}{\partial x_i \partial x_j} \right),
\]
where we use indicial notation with summation implied, $u_j$ is a velocity component, and $\tau$ is a characteristic flow time. We can estimate the flow time in terms of a characteristic speed $U$ and the total circulation distance: $\tau \sim (L_x + L_z)/U$. If there is no flow in the $y$ direction and no flow variation in the $x$ direction, we can estimate the gradients in the viscous damping term as well. Setting the result equal to the work (per unit volume) done by capillary forces and solving for $U$, we estimate a characteristic speed
\[
U \sim \frac{\Delta T \frac{\partial \sigma}{\partial T}}{2 \mu} \frac{L_z}{L_z+L_x}.
\]
As expected, the speed increases with $\Delta T$ and $\partial \sigma / \partial T$, which increase the thermocapillary force; and increases with $L_z$, which reduces viscous shear; but decreases with $\mu$, $L_z$, and $L_x$, which increases viscous drag. Again considering the model of~\cite{KoellnerBoeckSchumacher:2017}, we find $U \sim 0.4$~mm/s in the PbBi positive electrode and $U \sim 8$~mm/s in the LiCl-KCl electrolyte. Figure~19 in~\cite{KoellnerBoeckSchumacher:2017} shows velocities around 2~mm/s, so our velocity scaling argument seems to predict the correct order of magnitude. 

\begin{figure}
  \centering
  \includegraphics[width=\columnwidth]{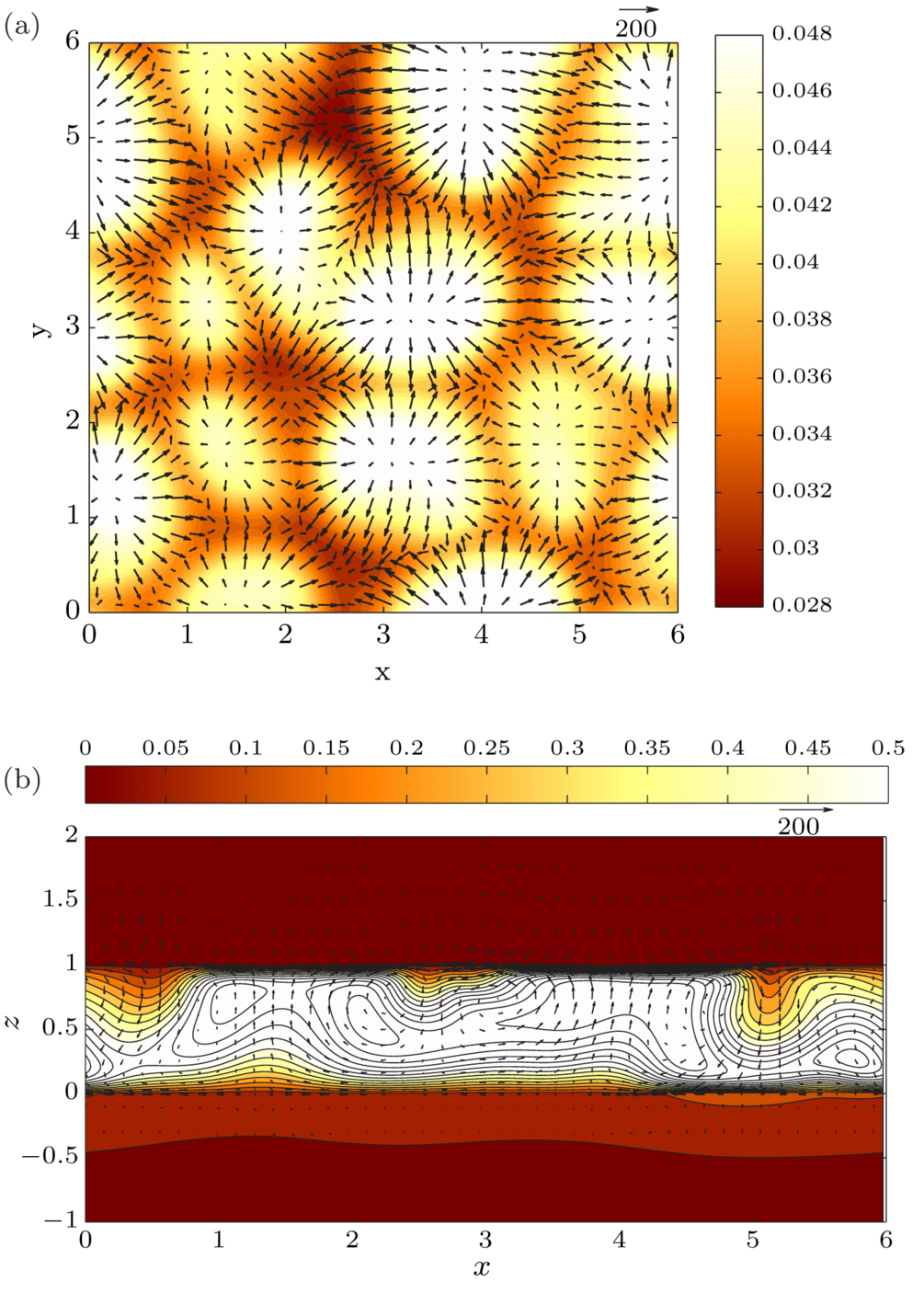}
  \caption{\label{fig:Koellner2017fig16ac}Marangoni flow in a three-layer liquid metal battery. Temperature is indicated in color, and velocity is indicated by arrows. The horizontal top surface of the electrolyte shows Marangoni cells (a), with downwellings where the temperature is highest. A vertical cross-section through the center of the battery (b) also shows downwellings, and indicates that the temperature is much higher in the electrolyte than in either electrode. Adapted from~\cite{KoellnerBoeckSchumacher:2017}, with permission.}
  \end{figure}

Simulations can give further insight into thermocapillary flow in liquid metal batteries. K\"ollner, Boeck, and Schumacher~\cite{KoellnerBoeckSchumacher:2017} considered a model liquid metal battery with uniform current density that caused Joule heating in all three layers, thereby causing both thermocapillary flow and buoyancy-driven thermal convection. As shown in Fig.~\ref{fig:Koellner2017fig16ac}, Marangoni cells are evident at the top of the electrolyte, and the temperature is much higher in the electrolyte than in either electrode, consistent with Fig.~\ref{fig:ShenZikanov2016fig8ad}. Using a range of layer thicknesses and current densities, the study found that five modes of thermal flow arise in typical liquid metal batteries. In order of decreasing typical speed, they are (1) thermal convection in the electrolyte, (2) thermal convection in the negative electrode, (3) thermocapillary flow driven by the top surface of the electrolyte, (4) thermocapillary flow driven by the bottom surface of the electrolyte, and (5) anti-convection~\cite{Welander:1964} in the positive electrode. The combined effects of buoyant and thermocapillary forces produce flows much like those produced by buoyancy alone, though thermocapillary forces slightly reduce the characteristic length scale of the flow. That observation is consistent with an earlier observation that thermocapillary and buoyant forces drive flows having different characteristic lengths~\cite{Koschmieder:1992}. In the electrolyte, thermocapillary forces always augment buoyant flow, but in the negative electrode, thermocapillary forces oppose and substantially damp buoyant flow when $Ma < 200$~\cite{KoellnerBoeckSchumacher:2017}. Electrolyte layers thinner than 2~mm exhibit neither thermocapillary flow nor thermal convection for realistic current densities (less than 2000~A/m\textsuperscript{2}). We raise one caveat: if the negative electrode is contained by a metal foam, flow there would likely be negligible. 

\subsection{Introduction to Solutal Marangoni flow}

Solutal Marangoni flow has been studied less than thermocapillary flow, and to our knowledge has not yet been addressed in the literature for the specific case of liquid metal batteries. One experimental and numerical study found a cellular flow structure reminiscent of the hexagons characteristic of the short-wavelength mode in thermocapillary flow~\cite{Koellner:2013}. A later experimental and numerical study by the same authors~\cite{Koellner:2015} varied the thickness of the fluid layer and its orientation with respect to gravity, finding that a two-dimensional simulation in which flow quantities are averaged across the layer thickness fails to match experiments with thick layers. The study also found that cells coarsen over time, perhaps scaling as $t^{1/2}$, where $t$ is time. 

Though studies of solutal Marangoni flow in liquid metal batteries have not yet been published, the phenomenon is likely, because charge and discharge alter the composition of the positive electrode. In past work, salt loss in lithium-chalcogen cells has been attributed to Marangoni flow~\cite{WalshGayArntzenEtAl:1973}. In the case where composition varies \emph{across} the interface, solutal Marangoni flow is possible only if material flows across the interface in the direction of increasing material diffusivity. In a liquid metal battery, the material of interest is the negative electrode material, e.g. Li, and the interface of interest is the one between molten salt electrolyte and liquid metal positive electrode. The diffusivity of Li in Bi is $1.2 \times 10^{-8}$~m\textsuperscript{2}/s, and the diffusivity of Li in LiBr-KBr has been calculated as $2.4 \times 10^{-9}$~m\textsuperscript{2}/s~\cite{Newhouse:2014}. A battery made with those materials would be prone to solutal Marangoni flow driven by composition varying across the interface during discharge, but not during charge. Solutal Marangoni flow driven by variations \emph{across} the interface is likely to occur in both short- and long-wavelength modes, depending on the appropriate Marangoni and Galileo numbers (analogous to eqs.~\ref{eq:Marangoni} and \ref{eq:Galileo}). However, a two-layer model for solutal Marangoni flow is unstable at any value of the Marangoni number~\cite{Davis:1987,Scriven:1964}. Variations of composition \emph{along} the interface will drive solutal Marangoni flow regardless of their values.

An estimate of the magnitude of solutal Marangoni flow would be useful. Even less is known about the rate of change of surface tension with composition than about the rate of change with temperature. Still, we can put an upper bound on the magnitude of solutal Marangoni flow, and compare to thermocapillary flow, by considering extreme cases. The force per unit length that drives thermocapillary flow is
\[
\left| \frac{\partial \sigma}{\partial T} \Delta T \right|.
\]
Again considering the same situation as~\cite{KoellnerBoeckSchumacher:2017}, we find a force per unit length around $1.8 \times 10^{-4}$~N/m. The force per unit length that drives solutal Marangoni flow is 
\[
\left| \frac{\partial \sigma}{\partial X} \Delta X \right|,
\]
Unfortunately, $\partial \sigma / \partial X$ is, to our knowledge, unknown in the literature for materials common to liquid metal batteries. Alternatively, we can consider the extreme case in which different regions of the interface are composed of different pure materials, so that the force per unit length is simply the difference between their (known) surface tensions. Using $\sigma_{PbBi} = 0.4086$~N/m at 500~K~\cite{NEA:2015}, $\sigma_{Li}=0.396$~N/m at 453~K~\cite{Davison:1968}, and $\sigma_{LiCl-KCl} = 0.122$~N/m at 823~K~\cite{Janz:1975}, we find $\sigma_{PbBi} - \sigma_{Li} = 1.3 \times 10^{-2}$~N/m and $\sigma_{Li}-\sigma_{LiCl-KCl} = 2.7 \times 10^{-1}$~N/m. These estimates are imprecise: considering temperature will change them by a few percent, and considering different battery chemistry will change them more. These estimate are also upper bounds. Nonetheless, these estimates are two to four orders of magnitude larger than the typical force per unit length that drives thermocapillary flow. If the true solutal forces reach even a few percent of these estimates, solutal Marangoni flow rivals or dominates thermocapillary flow in liquid metal batteries. Better constraints on the magnitude of solutal Marangoni flow~--- beginning with estimates of $\partial \sigma / \partial X$~---would be a valuable contribution for future work. 

\section{Interface instabilities}
\label{sec:interface}

\begin{figure}
  \centering
  \includegraphics[width=0.8\columnwidth]{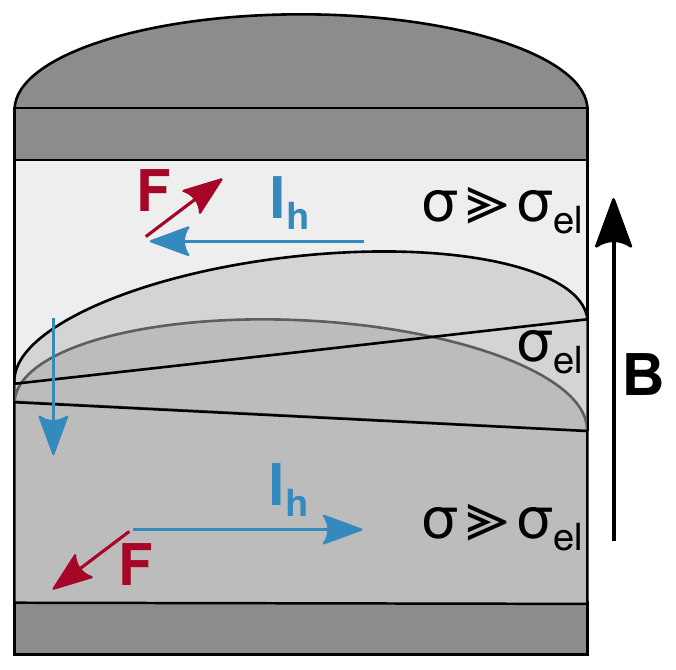}
  \caption{  \label{fig:lmb_sloshing}Sketch of a liquid metal cell undergoing an interfacial instability}
\end{figure}

\begin{figure}
  \centering
  \includegraphics[width=0.51\columnwidth]{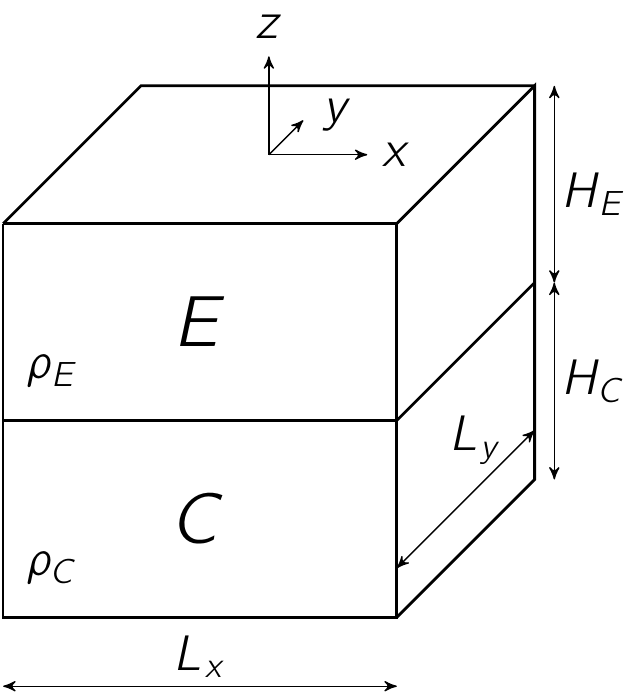}%
    \hfill%
  \includegraphics[width=0.44\columnwidth]{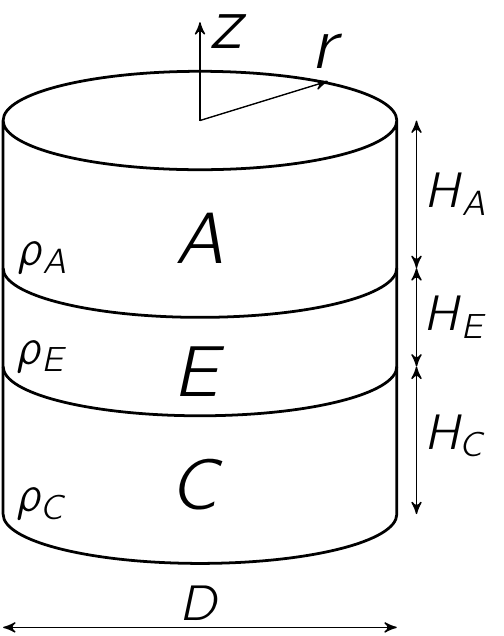}
  \caption{Characteristic dimensions and notations for an aluminium
    electrolysis cell (left) and a liquid metal battery (right).}
  \label{fig:AEC_LMB_dimensions}
\end{figure}

It is a well known phenomenon in Hall-H\'eroult, i.e., aluminium
electrolysis cells (AECs) that long wave instabilities can develop at the
interface of the cryolite and the liquid aluminium
\cite{Sele:1977,Davidson:2000,EvansZiegler:2007,MolokovElLukyanov:2011}.
Those instabilities are known as ``sloshing'' or ``metal pad roll
instability''. Not only because AECs gave the inspiration for the
initial LMB concept at MIT \cite{Bradwell:2006} it is worthwhile to
have a closer look at the role of interface instabilities in LMBs. If
the interface between a good electric conductor (metal,
$\sigma = O(10^6)$\,S/m) and a poor one (electrolyte,
$\sigma_{\text{el}} = O(10^2)$\,S/m) is slightly inclined with
respect to the horizontal plane, the current distribution inside the layers 
changes. In the metal layer(s), horizontal perturbation currents
($I_{\text{h}}$) arise as sketched in
Fig.~\ref{fig:lmb_sloshing}. Those horizontal currents interact with
the vertical component of a background magnetic field generated, e.g.,
by the current supply lines, generating Lorentz forces that set the
metal layer into motion. This mechanism was first explained by Sele
\cite{Sele:1977} for AECs. As a consequence, gravity waves with a
characteristic length of the vertical cell size develop and
culminate in a sloshing motion of the aluminium. Wave amplitudes may
become large enough to reach the graphite negative electrodes and 
short-circuit the cell, thereby terminating the reduction
process. In order to prevent the waves from contacting the negative electrodes for
a cell current of about 350\,kA, considered as an upper limit for
modern cells \cite{OeyeMasonPetersonEtAl:1999}, a cryolite layer
at least 4.5\,cm thick is required~\cite{Davidson:2000}. These
boundary conditions mean that nearly half of the cell voltage is spent
overcoming the electrolyte resistance, and the corresponding
electric energy is converted to heat \cite{Davidson:2000}. Reducing
the electrolyte layer thickness by even a few millimeters would result in
large cost savings, but is made impossible by the sloshing
instabilities. Admittedly, Joule heating is not entirely wasted, because it maintains the high cell temperature and to permits the
strong wall cooling that allows the formation of the side-wall
protecting ledge \cite{EvansZiegler:2007}. Metal pad
rolling in AECs, which typically have a rectangular cross-section, occurs if the parameter
\begin{equation}
  \beta = \frac{J B_{z}}{g \Delta\rho_{\text{CE}}} \cdot \frac{L_x}{H_{\text{E}}} \cdot \frac{L_y}{H_{\text{C}}}
  \label{eq:beta_Sele}
\end{equation}
exceeds a critical value $\beta_{\text{cr}}$. Here $J$ and $B_{z}$
denote the absolute values of the cell's current density and of the
vertical component of the background magnetic field, respectively, $\Delta \rho_{\text{CE}}$ is the density difference between
cryolite and aluminium, and $H_{\text{E}}$, $H_{\text{C}}$, $L_x$,
$L_y$ refer to the layer heights and the lateral dimensions of the
AEC, respectively. See Fig.~\ref{fig:AEC_LMB_dimensions} (left) for reference.
The first factor in Eq.~\eqref{eq:beta_Sele} is the ratio
of Lorentz force to gravity force, and the latter ones are ratios
of layer height to lateral cell dimension.

Bojarevi{\v{c}}s and Romerio \cite{BojarevicsRomerio:1994} obtained an
expression for $\beta_{\text{cr}}$ depending on wave numbers of
gravity waves $m, n$ in $x, y$ direction developing in rectangular
cells:
\begin{equation}
  \label{eq:Bojarevics_Romerio}
  \beta_{\text{cr}} = \pi^2 \left| m^2\frac{L_y}{L_x} - n^2\frac{L_x}{L_y}\right|.
\end{equation}
According to Eq.~\eqref{eq:Bojarevics_Romerio} cells with square or
circular cross-section are always unstable because their lateral
dimensions are equal and thus $\beta_{\text{cr}}=0$. Davidson and
Lindsay \cite{DavidsonLindsay:1998} came to a similar conclusion
regarding the instability threshold for circular and square cells
using both shallow water theory and a mechanical analogue. 

It can be expected that three-layer systems like Al refinement
cells (cf. \S\ref{sec:history-refinement-cells}) and LMBs will
exhibit features similar to those found in AECs, but the addition of
the second electrolyte-metal interface should enrich system
dynamics. Knowledge on three-layer systems bearing interface normal
currents is currently relatively scarce. Sneyd \cite{Sneyd:1985}
treated the case while modeling an electric-arc furnace, assuming a
density of zero for the upper phase and semi-infinite upper and lower
layers. He took only the azimuthal magnetic field produced by the
current into account and did not consider the action of an additional
background field. In addition to long wave instabilities Sneyd
predicted short wave instabilities of both sausage and kink
type. To the best of our knowledge, experimental
results on current-driven interface instabilities in three-layer systems have
not been reported to date. The cause of the violent motions reported
by several authors
\cite{Frary:1925,EdwardsFraryJeffries:1930,PearsonPhillips:1957,Zeerleder:1955},
and already mentioned in \S\ref{sec:history-refinement-cells}, is
uncertain. Frary \cite{Frary:1925} describes the motion as swirling and
attributes it to the interaction of the vertical current within the cell and the
magnetic fields of the horizontal current leads. To prevent
inter-mixing of the negative and positive electrodes, the electrolyte layer
has to be as thick as 25\,cm. This is costly in terms of
energy but tolerable if the cell is operated as an electrolytic cell.
In galvanic mode, open circuit voltage (OCV) obviously limits the
permissible current and the resistance of thick electrolyte layers is
prohibitive. For LMBs to have an acceptable voltage efficiency the
electrolyte thickness must not exceed a few millimeters, so maintaining interface stability is more difficult. 

Zikanov \cite{Zikanov:2015} was the first to discuss sloshing behavior
in three-layer systems explicitly addressing LMBs. He used a
mechanical analogue inspired by Davidson and Lindsay's
\cite{DavidsonLindsay:1998} movable aluminium plate model for the
basic features of sloshing in AECs. Instead of one
plate mimicking the aluminium layer of an AEC, Zikanov assumed two
slabs of solid metals to be suspended as pendula above and below the
electrolyte layer \cite{ZikanovShen:2016}. This model replaces the
hydrodynamic problem by a system possessing only four degrees of
freedom associated with the two-dimensional oscillations of each
pendulum. The Lorentz force due to the interaction of the vertical
background magnetic field and the horizontal currents can cause an
instability if

\begin{equation}
  C_{\text{A}} \frac{J B_{z}  L_x^2}{12 g \rho_{\text{A}} H_{\text{E}} H_{\text{A}}} + C_{\text{C}} \frac{J B_{z} L_x^2}{12 g \rho_{\text{C}} H_{\text{E}} H_{\text{C}}}
  > \left|  1 - \frac{\omega_x^2}{\omega_y^2} \right| .
  \label{eq:ZikanovShen_2016_Bz}
\end{equation}

Here $C_{\text{A}}$ and $C_{\text{C}}$ are constants of order one that account for geometry~\cite{Zikanov:2015}, and $\rho_{\text{A}}$ and
$\rho_{\text{C}}$ denote the densities of the negative and positive electrodes,
respectively. The pendula oscillate with their natural gravitational
frequencies $\omega_x$ and $\omega_y$. Again, it is evident from
Eq.~\eqref{eq:ZikanovShen_2016_Bz} that circular or square
cross-sections are predicted to be always unstable.

Zikanov \cite{Zikanov:2015,ZikanovShen:2016} discussed an additional
instability that may arise even in the absence of a background
magnetic field due to the interaction of $J$-generated  azimuthal
magnetic field $B_{\varphi}$ with the current perturbations. He
finds the system to be unstable if 

\begin{equation}
  \frac{\mu_0 J^2 D^2}{64 g} \left(
    \frac{D^2}{12 \rho_{\text{A}} H_{\text{A}} H^2_{\text{E}}v} +
    \frac{D^2}{12 \rho_{\text{C}} H_{\text{C}} H^2_{\text{E}}} +
    \frac{1}{\rho_{\text{A}} H_{\text{E}}} -
    \frac{1}{\rho_{\text{C}} H_{\text{E}}}
    \right) > 1 .
  \label{eq:ZikanovShen_2016_Bphi}
\end{equation}

As estimated by Zikanov \cite{Zikanov:2015}, for rectangular cells the
instability due to the interaction of the perturbation currents with
the azimuthal field of the main current described by criterion \eqref{eq:ZikanovShen_2016_Bphi}
appears to be more dangerous than that caused by the action of the
background magnetic field on the horizontal compensating currents
\eqref{eq:ZikanovShen_2016_Bz}.

It should be mentioned that criteria predicting instability onset for
any non-vanishing Lorentz force neglect dissipative effects caused by
magnetic induction and viscosity as well as the influence of surface
tension \cite{WeberBecksteinHerremanEtAl:2017}.

Weber et al.~\cite{WeberBecksteinGalindoEtAl:2017, WeberBecksteinHerremanEtAl:2017} investigated the
metal pad roll instability in an LMB using a volume-of-fluid method
adapted from the finite volume code OpenFOAM
\cite{WellerTaborJasakFureby:1998} and supplemented by electromagnetic
field calculations to solve the full Navier-Stokes equations. The
material properties correspond to the special case of the
Mg$|$KCl-MgCl\textsubscript{2}-NaCl$|$Sb system (see
Table~\ref{tab:LMB_properties} for an overview of typical systems).

\begin{figure}
  \centering \includegraphics[width=\columnwidth]{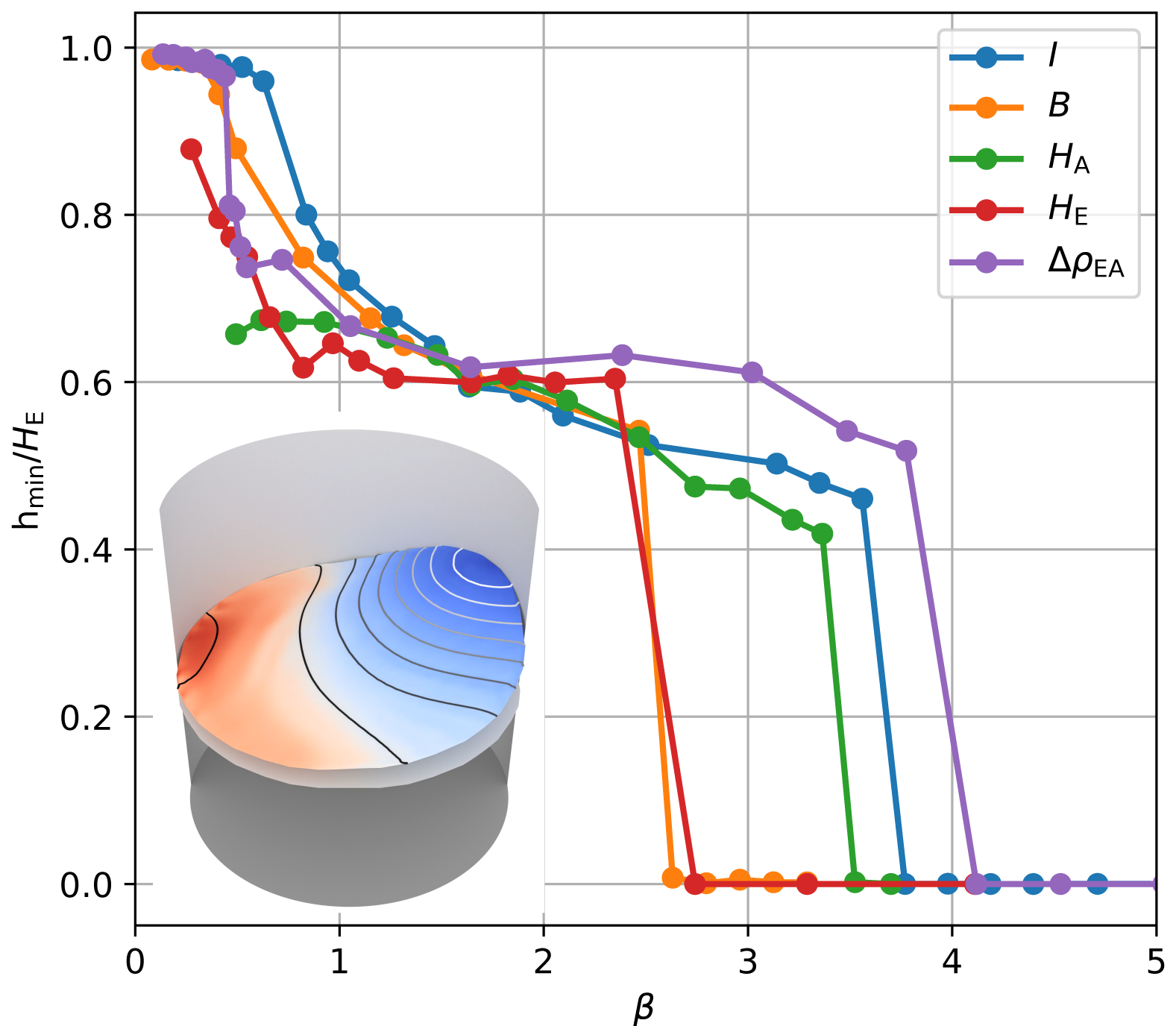}
  \caption{Minimum electrolyte layer height $h_{\text{min}}$ depending
    on $\beta$ according to Eq.~\eqref{eq:WeberEtAl_beta_krit} for the
    Mg$|$KCl-MgCl\textsubscript{2}-NaCl$|$Sb system adapted from Weber
    et al.~\cite{WeberBecksteinHerremanEtAl:2017}. For each curve,
    only the parameter named in the legend is varied, the other ones
    stay constant ($j = 1$\,A/cm\textsuperscript{2}, $B_{z} = 10$\,mT,
    $H_{\text{A}} = 4.5$\,cm, $H_{\text{E}} = 1$\,cm,
    $\rho_{\text{A}} = 1577$\,kg/m\textsuperscript{3}).
    $\Delta \rho_{\text{EA}} = \rho_{\text{E}} - \rho_{\text{A}}$, the
    inset shows a snapshot of the anode/electrolyte interface for
    $\beta=2.5$.}
  \label{fig:beta_Norbert_rescaled}
\end{figure}

As expected, if one interface is set in motion and the other
remains nearly at rest, a criterion similar to the Sele criterion
\eqref{eq:beta_Sele} can be formulated:
\begin{equation}
  \beta = \frac{J B_{z} D^2}{g (\rho_{\text{E}} - \rho_{\text{A}}) H_{\text{A}} H_{\text{E}}} > \beta_{\text{cr}},
  \label{eq:WeberEtAl_beta_krit}
\end{equation}
using the density differences between negative electrode and electrolyte and
the respective layer heights. Sloshing in circular cells sets in above
a relatively well defined $\beta_{\text{cr, sloshing}} =
0.44$. Short-circuiting needs more intense Lorentz forces and happens
above $\beta_{\text{cr, short-circuit}} \approx
2.5$. The validity of both values is limited to the Mg$||$Sb system
and to the aspect ratios $H_{\text{A}}/D =
0.45$ and $H_{\text{E}} = 0.1$ investigated by Weber et
al.~\cite{WeberBecksteinHerremanEtAl:2017}, see Fig.~\ref{fig:beta_Norbert_rescaled}. 

Bojarevi{\v{c}}s et
al.~\cite{BojarevicsTucsPericleous:2016,BojarevicsTucs:2017}
numerically investigated the Mg$||$Sb system as well, but used a
shallow water approximation combined with the electromagnetic field
equations. They considered a cell with a 8 $\times$
3.6\,m\textsuperscript{2} cross-section and Mg and Sb layers both
20\,cm in height divided by a 5 or 8\,cm thick electrolyte. In
agreement with the results of Weber et
al.~\cite{WeberBecksteinGalindoEtAl:2017,WeberBecksteinHerremanEtAl:2017},
Bojarevi{\v{c}}s et
al.~\cite{BojarevicsTucsPericleous:2016,BojarevicsTucs:2017} found the
interface between negative electrode and electrolyte to be much more
sensitive to the instability than the interface between the electrolyte and positive electrode. The difference is explained by density differentials: the electrolyte typically has density closer to that of the negative electrode than the positive electrode.  Bojarevi{\v{c}}s and Tucs \cite{BojarevicsTucs:2017}
further optimized the magnetic field distribution around the LMB by
re-using a commercial Trimet 180\,kA cell series in their
simulation. While the unoptimized cell could only be stabilized for
the 8\,cm thick electrolyte, the optimized cell was able to operate
with 2.5\,cm electrolyte height. However, even in the latter case the
voltage drop in the electrolyte is still found to be excessive with
0.49\,V at a current of 100\,kA.

\begin{table*}
  \caption{Coupling parameter $\mathcal{A}_{\text{g}}$ calculated for
    different possible working material combinations. The densities
    are reported at working temperature $T_{\text{op}}$, adapted from \cite{HorstmannWeberWeier:2017}} \label{tab:LMB_properties}
  \centering
  \begin{tabular}{l@{\hspace{0.3em}}llcrrrrr}
    & &  Electrodes & Electrolyte & $T_{\text{op}}$  & $\rho_{\text{A}}$  & $\rho_{\text{E}}$ & $\rho_{\text{C}}$ & $\mathcal{A}_{\text{g}}$ \\
    & &             &                 & ($^{\circ}$C) & \multicolumn{3}{c}{(kg m\textsuperscript{-3})} & \\\hline
    \multirow{3}{*}{\rotatebox{90}{strongly}} &
                                                \multirow{3}{*}{\rotatebox{90}{coupled}} 
      & Li$||$Se & LiCl-LiF-LiI      & 375 & 497 & 2690 & 3814 & 0.51\\ 
    & & Al$||$Al-Cu$^{*}$ & AlF$_3$-NaF-CaCl$_2$-NaCl & 800 & 2300 & 2700 & 3140 & 1.1\\	
    & & Li$||$Te & LiCl-LiF-LiI      & 475 & 489 & 2690 & 5782 & 1.41\\ \hline
    \multirow{4}{*}{\rotatebox{90}{weakly}} &
                                              \multirow{4}{*}{\rotatebox{90}{coupled}} 
      & Na$||$Sn & NaCl-NaI          & 625 & 801 & 2420 & 6740 & 2.67\\
    & & Li$||$Bi & LiCl-LiF-LiI      & 485 & 488 & 2690 & 9800 & 3.22\\
    & & Na$||$Bi & NaCl-NaI-NaF      & 550 & 831 & 2549 & 9720 & 4.18\\
    & & K$||$Hg  & KBr-KI-KOH & 250  & 640 & 2400 & 12992 & 6.02\\ \hline
    \multirow{3}{*}{\rotatebox{90}{not}} &
                                           \multirow{3}{*}{\rotatebox{90}{coupled}}  
      & Ca$||$Sb & CaCl$_2$-LiCl     & 700 & 1401 & 1742 & 6270 & 13.28\\
    & & Ca$||$Bi & CaCl$_2$-LiCl     & 550 & 1434 & 1803 & 9720 & 21.43\\
    & & Mg$||$Sb & KCl-MgCl$_2$-NaCl & 700 & 1577 & 1715 & 6270 & 33.06\\
  \end{tabular}
\end{table*}

Horstmann et al.~\cite{HorstmannWeberWeier:2017} investigated the wave
coupling dynamics of both interfaces by applying potential theory as
well as direct numerical simulations to LMBs with circular
cross-section.  While interface tension should be taken into account
for (very) small cells and large wave numbers, it is negligible in the
limit of large-scale LMBs. There, the waves are purely gravitational
ones and the strength of their coupling depends only on the ratio of
the density differences

\begin{equation}
\mathcal{A}_{\text{g}} = \frac{\rho_{\text{C}} - \rho_{\text{E}}}{\rho_{\text{E}} - \rho_{\text{A}}} .
  \label{eq:Horstmann_A_gravity}
\end{equation}

Thus, for practical cases, $\mathcal{A}_{\text{g}}$ is the control
parameter that determines how strongly both interfaces interact. Wave
onset is described by Sele-like parameters extended by interface
tension terms for both interfaces. The expressions reduces to the Sele
criterion \eqref{eq:WeberEtAl_beta_krit} in the limit of large LMBs
considered here.

At the same time $\mathcal{A}_{\text{g}}$ describes for
thin electrolyte layers ($H_{\text{E}} \rightarrow 0$)
the ratios of amplitudes and frequencies of the (anti-symmetric) waves

\begin{equation}
  \left| \frac{\hat \eta^{mn}_{\text{AE}}}{\hat \eta^{mn}_{\text{EC}}} \right| =
  \frac{\omega^2_{\text{EC}}}{\omega^2_{\text{AE}}} =
  \mathcal{A}_{\text{g}} .
  \label{eq:Horstmann_A_waves}
\end{equation}
Here $\hat \eta^{mn}_{\text{AE}}$, $\hat \eta^{mn}_{\text{EC}}$ denote the amplitudes
of the of the waves at the AE and EC interfaces, respectively, with the
azimuthal wave number $m$ and radial wave number $n$. $\omega_{\text{AE}}$ and
$\omega_{\text{EC}}$ are the corresponding frequencies.

The waves at both interfaces can be considered as coupled in the range
$0.1 < \mathcal{A}_{\text{g}} < 10$. If the metal layer
with density more similar to the electrolyte is thinner, the limiting values have to be corrected by the metal
layer height ratio $H_{\text{C}}/H_{\text{A}}$. The coupled regime can be further divided into ``weakly coupled''
( $0.1 < \mathcal{A}_{\text{g}} \lesssim 0.7$,
$2 \lesssim \mathcal{A}_{\text{g}} <10$) and ``strongly
coupled'' ($0.7 \lesssim \mathcal{A}_{\text{g}} \lesssim
2$) regimes. The threshold values are empirical. In the weakly coupled regime
the interfaces are anti-symmetrically displaced and co-rotate in the
direction determined by the more prominent wave. Dynamics in the
strongly coupled regime are more complex. For moderate
$\beta_{\text{AE}} \approx 1.6$ both metals rotate in opposite
directions deforming the electrolyte layer into a bulge (``bulge
instability''). Higher $\beta_{\text{AE}}$ ($\approx 3.2$)
leads to synchronously rotating metal
pads (``synchronous tilting instability''). These strongly coupled instabilities may not occur in cells that are not circular, however.

While the two strongly coupled LMB types (Li$||$Te and Li$||$Se) have
limited practical relevance due to the scarcity of their positive electrode
materials, three-layer refinement cells are almost always strongly coupled.
Gesing et al \cite{GesingDasLoutfy:2016, GesingDas:2017} formulate it
as a characteristic of their Mg-electrorefinement method that the
electrolyte has to have a density halfway between that of Al and Mg,
i.e., they require a coupling parameter $\mathcal{A}_{\text{g}} = 1$.

Zikanov \cite{Zikanov:2017} used the St.~Venant shallow water
equations complemented by electromagnetic force terms to model the
rolling pad instability in LMBs with rectangular cross-sections. In
accordance with Horstmann et al.~\cite{HorstmannWeberWeier:2017},
Zikanov \cite{Zikanov:2017} found that the wave dynamics depend on the
ratio of the density jumps at both interfaces. If the density jump at
one interface is much smaller than at the other, only the former
develops waves, and the situation is very similar to that in AECs. In
particular the influence of the horizontal aspect ratio $L_x/L_y$ on
the critical value of the Sele criterion is quite strong and resembles
the situation in ARCs. This strong effect can be explained by the fact
that the aspect ratio determines the set of available natural
gravitational wave modes and the strength of the electromagnetic field
that is needed to transform them into a pair with complex-conjugate
eigenvalues \cite{Zikanov:2017,DavidsonLindsay:1998}.

For comparable density jumps at the interfaces, Zikanov's
\cite{Zikanov:2017} results again agree with those of Horstmann et
al.~\cite{HorstmannWeberWeier:2017} in that both interfaces are
significantly deformed. The system behavior becomes more complex and
is different from that found in AECs. The waves of both interfaces can
couple either symmetrically or anti-symmetrically. Zikanov
\cite{Zikanov:2017} found examples where the presence of the second
interface stabilizes the system, which was not predicted by his
two-slab model \cite{Zikanov:2015}, whose simplifications are probably
too strong to capture this part of the dynamics.

\section{Tayler Instability}
\label{sec:Tayler}

\begin{figure}
  \centering
  \includegraphics[width=0.8\columnwidth]{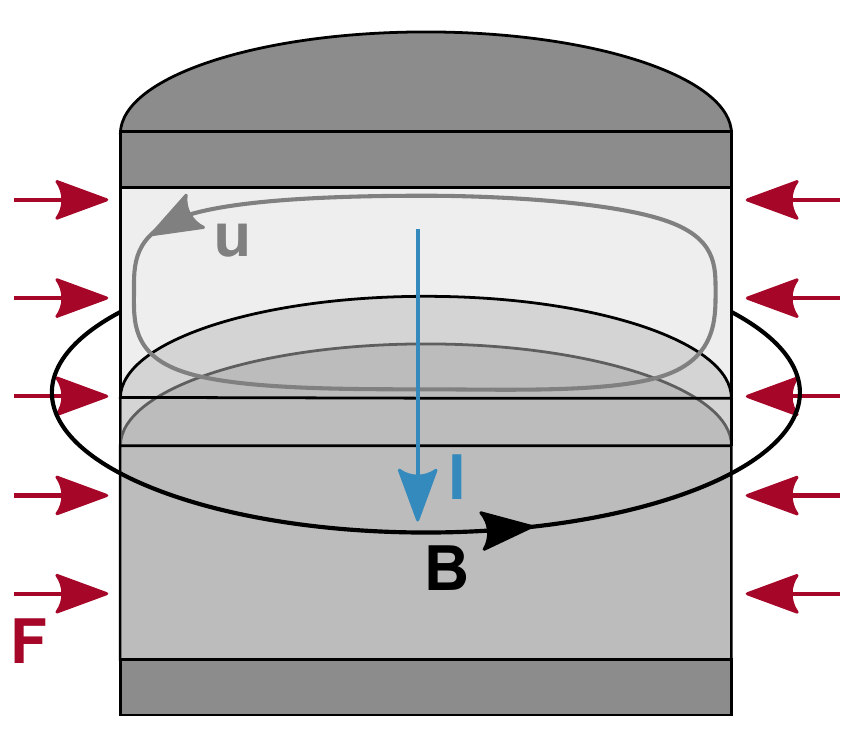}  
  \caption{Sketch of a liquid metal cell susceptible to the Tayler instability.}
  \label{fig:lmb_tayler}
\end{figure}

Electric currents induce magnetic fields and interact with those fields, sometimes bringing unexpected consequences. Suppose a current runs axially and has azimuthally symmetric current density $\bm{J}$, as sketched in Fig.~\ref{fig:lmb_tayler}. Then, by the right-hand rule, it induces a magnetic field $\bm{B}_\varphi$ that is purely in the azimuthal direction, and interacts with that field to cause a Lorentz force per unit volume $\bm{F_L} = \bm{J} \times \bm{B}_\varphi$ directed radially inward. That force can be understood as a magnetic pressure. If the current flows through a fluid that is incompressible, we might expect the magnetic pressure to have no effect. However, Tayler~\cite{Tayler:1957,Tayler:1973} and Vandakurov~\cite{Vandakurov:1972} showed that if the fluid is inviscid ($\nu=0$) and a perfect conductor ($\sigma_E=\infty$), and the induced magnetic field satisfies
\begin{equation}
\label{eq:TaylerOnset}
\frac{\partial}{\partial r}\left( r B_\varphi^2 \right) < 0,
\end{equation}
then the stagnant system is unstable. Given an infinitesimal perturbation, the current drives fluid flow, initially with azimuthal wave number $m=1$. That phenomenon, known as the Tayler instability in astrophysics and as the kink instability in plasma physics, must also be considered for liquid metal batteries, which support large axial currents. 

In stars, the Tayler instability can in theory overcome gravitational stratification to drive dynamo action, and has therefore been proposed as a source of stellar magnetic fields~\cite{Spruit:2002}. In fusion plasma devices, the kink instability~\cite{Rosenbluth:1973} can disrupt the magnetic fields that prevent plasma from escaping and must therefore be avoided. First encountered in z-pinch experiments in the 1950s, the kink instability has been studied extensively and reviewed in the plasma physics literature (e.g.,~\cite{Freidberg:1982}). 

Soon after liquid metal batteries were proposed for grid-scale storage, Stefani \emph{et al.}~\cite{StefaniWeierGundrumEtAl:2011} observed that the technology would be susceptible to the Tayler instability, and that if the resulting flow were strong enough, it could cause rupture the electrolyte layer, destroying the battery. Their observation prompted a series of studies considering methods to avoid the Tayler instability in liquid metal batteries and the likelihood of it causing rupture. 

The Tayler instability can be avoided or damped using a variety of
techniques. First, in a real fluid with nonzero viscosity and
imperfect electrical conductivity, the onset criterion given by
Eq.~\ref{eq:TaylerOnset} is no longer strictly correct, because
viscosity and resistance damp the instability, so that it occurs only
if the total current (or Hartmann number) exceeds a nonzero critical
value. Second, the instability can be avoided by cleverly routing the
battery current to prevent the condition expressed by
Eq.~\ref{eq:TaylerOnset}. Instead of building a cylindrical liquid
metal battery carrying an axial current, one can build a battery that
is a cylindrical annulus, with an empty central bore. Carrying no
\begin{figure}
  \centering
  \includegraphics[width=\columnwidth]{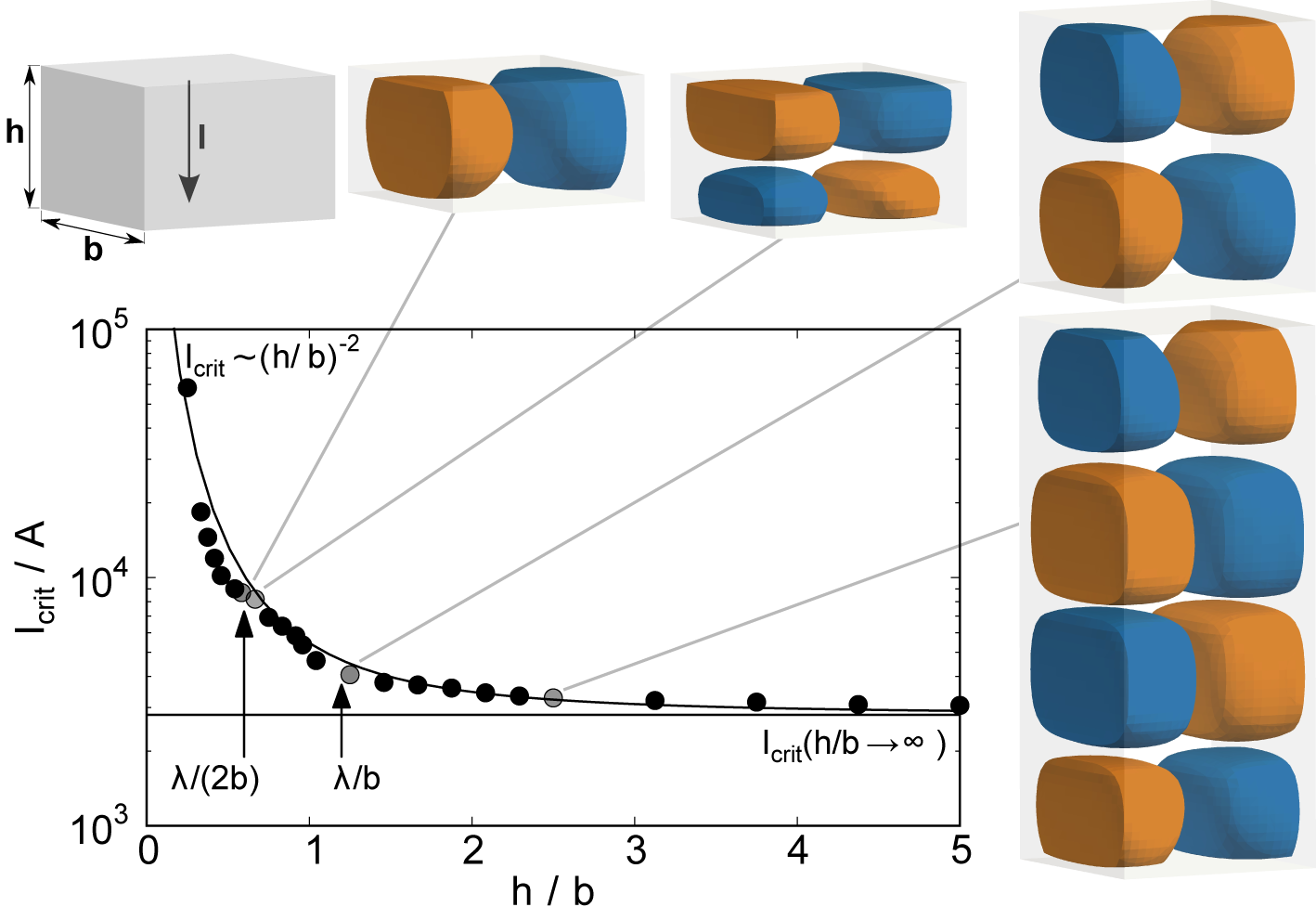}
  \caption{Critical current for the Tayler instability depending on
    the aspect ratio of a cuboid with square cross section filled with
    Na. The insets show contours of the induced vertical magnetic
    field component. Adapted from \cite{WeberGalindoStefaniEtAl:2013}.}
  \label{fig:TI_I_crit_vs_aspect_ratio}
\end{figure}
current, the bore does not contribute to the induced magnetic field
$\bm{B}_\varphi$, so that a larger current is required before
onset~\cite{StefaniWeierGundrumEtAl:2011}. Even better is to route the
battery current back through the bore in the \emph{opposite}
direction, which prevents the Tayler instability altogether, since
Eq.~\ref{eq:TaylerOnset} is never
satisfied~\cite{StefaniWeierGundrumEtAl:2011,WeberGalindoStefaniWeier:2014}. However,
ohmic losses in the wire would reduce the
available voltage. Third, shearing the fluid azimuthally can damp the
Tayler instability~\cite{Shumlak:1995}, though imposing shear is more
practical in plasma devices than in liquid metal batteries. Fourth,
imposing external magnetic fields~--- either axial or transverse~---
damps the Tayler instability~\cite{WeberGalindoStefaniWeier:2014}, in
a process often compared to the damping of thermal convection by vertical magnetic fields~\cite{Chandrasekhar:1954,Burr:2001}. Finally, rotation also damps the Tayler instability~\cite{WeberGalindoStefaniEtAl:2013}, though imposing rotation may be impractical for liquid metal batteries. 

A variety of engineering solutions to avoid or damp the Tayler instability in liquid metal batteries are now known. The characteristics of the Tayler instability, and the particular situations in which it arises, have also been studied extensively in recent work. The instability was observed directly in a laboratory experiment: when the axial current applied to a cylindrical volume of GaInSn alloy exceeded a critical value, the induced axial magnetic field was observed to grow as the square root of the current~\cite{SeilmayerStefaniGundrumEtAl:2012}. Below the critical current, which nearly matched earlier numerical predictions~\cite{Rudiger:2011}, no axial field was induced. A central bore was added to the vessel, and as expected, the critical current grew with bore size, ranging from about 2500~A to about 6000~A. Flow due to the Tayler instability was found to compete with thermal convection caused by Joule heating (see \S\ref{sec:convection}). 

A series of numerical studies have also considered the Tayler
instability, with the benefit of being able to characterize its
properties in detail. Initial simulations with purely axial current
and perfectly conducting boundaries found the Tayler instability to
occur at $Ha \sim
20$~\cite{WeberGalindoStefaniEtAl:2013,WeberGalindoStefaniWeier:2015,StefaniGalindoKasprzykEtAl:2016}. (The
Hartmann number (see Eq.~\ref{eq:Hartmann}) is known to control
onset~\cite{Rudiger:2011}.) Using parameters similar to
experiments~\cite{SeilmayerStefaniGundrumEtAl:2012}, the simulations
found similar critical currents. The simulations also demonstrated the
importance of aspect ratio: liquid metal layers with high aspect ratio
(narrow and tall) are more susceptible to the Tayler
instability~\cite{StefaniGalindoKasprzykEtAl:2016}, see
Fig.~\ref{fig:TI_I_crit_vs_aspect_ratio}. Simulations have also shown
the effectiveness of engineering solutions for avoiding or damping the
Tayler instability, including adding a central bore, routing current
through the bore in the opposite direction, and applying axial or
transverse magnetic fields~\cite{WeberGalindoStefaniWeier:2014}. The
Tayler instability breaks chiral symmetry during its
growth~\cite{WeberGalindoStefaniWeier:2015} and might therefore
provide a link between planetary tides and the solar
cycle~\cite{StefaniGieseckeWeberWeier:2016}. By allowing for current
that is not purely axial, and boundaries that are not perfect
conductors, a later simulation~\cite{WeberGalindoPriedeEtAl:2015}
incorporated more realistic current collectors and considered their
effects. Current collectors with lower conductivity damp the Tayler
instability, as do fluids with lower
conductivity~\cite{WeberGalindoPriedeEtAl:2015}. That study also
found, however, that electro-vortex flows (see
\S\ref{sec:electro-vortex}) may play a larger role in the fluid
mechanics of liquid metal batteries than the Tayler instability.
A numerical linear stability analysis of the Tayler instability found critical currents consistent with prior work and described the instability as an edge effect, governed by curvature of the magnetic field~\cite{Priede:2016}. Finally, another simulation study found that the Tayler instability occurs in Mg-based liquid metal batteries with current density $J=300$~mA/cm\textsuperscript{2} (a value typical for batteries being commercialized) if the battery radius exceeds 0.43~m, and causes rupture if the radius exceeds 3~m~\cite{HerremanNoreCappaneraGuermond:2015}. Though there are economic advantages to increasing battery size, we are unaware of prototypes that large. 

To summarize: The Tayler instability is a magnetohydrodynamic phenomenon that drives flow when a large axial current passes through liquid metal. Because of the magnitude of the currents involved, and because today's liquid metal battery designs have large aspect ratio, the Tayler instability may not yet affect the technology. For larger, next-generation batteries, however, the Tayler instability should be considered carefully. A variety of engineering solutions to avoid or damp the instability are now known. 

\section{Electro-Vortex Flow}
\label{sec:electro-vortex}

The Tayler instability discussed in \S\ref{sec:Tayler} is an interaction of an electrical current with the magnetic field produced by the current itself. If condition~\ref{eq:TaylerOnset} is satisfied and enough current runs, the instability drives flow even in the case of a purely axial, azimuthally symmetric current. Currents of other shapes can also interact with their own magnetic fields to produce forces. In particular, if the Lorentz force produced by self-interaction has nonzero curl, a conductive fluid flows even with arbitrarily weak currents in the absence of perturbations. The phenomenon, known as ``electro-vortex flow'', has been reviewed in detail~\cite{Bojarevics:1989} and is commonly studied in the context of vacuum arc remelting~\cite{Davidson:1999,Davidson:2001}. Here we will give a brief overview and consider the implications of electro-vortex flow for liquid metal batteries. 

To show that a Lorentz force with nonzero curl drives flow, we consider the conservation of momentum of an incompressible Newtonian fluid acted upon by a force per unit mass $\bm{F}$:
\[
\frac{\partial \bm{u}}{\partial t} + \left( \bm{u} \cdot \bm{\nabla} \right) \bm{u} = -\frac{1}{\rho} \bm{\nabla}p + \nu \nabla^2 \bm{u} + \bm{F}.
\]
Here $\bm{u}$ is the fluid velocity. It follows that stagnant fluid ($\bm{u}=0$) conserves momentum if and only if 
\begin{equation}
\label{eq:pressureForce}
\frac{1}{\rho} \bm{\nabla} p = \bm{F}.
\end{equation}
One example, of course, is the gravitational force $\bm{F}=-g \hat{\bm{z}}$, where we take $\hat{\bm{z}}$ as vertical. In that case, Eq.~\ref{eq:pressureForce} may be integrated directly, yielding the familiar fact that the hydrostatic pressure a distance $h$ below the fluid surface is $p=\rho g h$. On the other hand, if $\bm{\nabla} \times \bm{F} \neq 0$, Eq.~\ref{eq:pressureForce} has no solution, since the curl of a gradient is always zero. In that case, we conclude that our assumption $\bm{u}=0$ was false, that is, if a fluid is acted upon by a force $\bm{F}$ for which $\bm{\nabla} \times \bm{F} \neq 0$, the fluid must flow. 

If we are interested in the Lorentz force per unit volume $\bm{F} = \bm{J} \times \bm{B}$ due to interaction of an electrical current with its own magnetic field, and under the simplifying assumption that all quantities are azimuthally symmetric, it can be shown~\cite{Bojarevics:1989} that $\bm{\nabla} \times \bm{F} \neq 0$ simplifies to 
\[
\frac{\partial B_\varphi}{\partial z} \neq 0. 
\]
The magnetic field $B_\varphi$ depends, via the Biot-Savart law, on
the current density $\bm{J}$. So in azimuthally symmetric systems like
cylindrical liquid metal batteries, electro-vortex flow occurs when
$\bm{J}$ varies axially. In fact, it can be shown that any divergent
current density in an axisymmetric system causes electro-vortex
flow. One canonical example is flow driven by a current from the
center of a hemisphere to the hemisphere's surface~\cite{Bojarevics:1989}. Taking the additional assumption that the resulting flow is irrotational allows analytical solution of many other cases as well. However, many are similarity solutions for which boundary conditions are evaluated at infinity, hindering their application to technological applications like liquid metal batteries. 

The details of electro-vortex flow are typically studied via simulation or experiment. Flows typically converge where the current density is highest, causing the largest magnetic pressure. In azimuthally symmetric situations, the result is a poloidal circulation. Current density can be shaped by choosing electrode geometry~\cite{Kolesnichenko:2002} or by adding a nearby ferromagnetic object, which concentrates magnetic field lines~\cite{Kolesnichenko:2005}. Electro-vortex flow can cause substantial pressure and was therefore recognized as a promising mechanism for pumping liquid metals in technological applications including metals processing and nuclear cooling~\cite{Denisov:1999}. Subsequent efforts produced pumps that drive liquid metal in flat channels with corners~\cite{Khripchenko:2008}, through Y- and $\Psi$-shaped junctions~\cite{Denisov:2012}, in a winding-free pump composed of a pair of junctions~\cite{Dolgikh:2014}, and centrifugally~\cite{Denisov:2016}. Electro-vortex flow also occurs in cylindrical steel furnaces~\cite{Kazak:2011}. 

\begin{figure}
  \centering
  \includegraphics[width=\columnwidth]{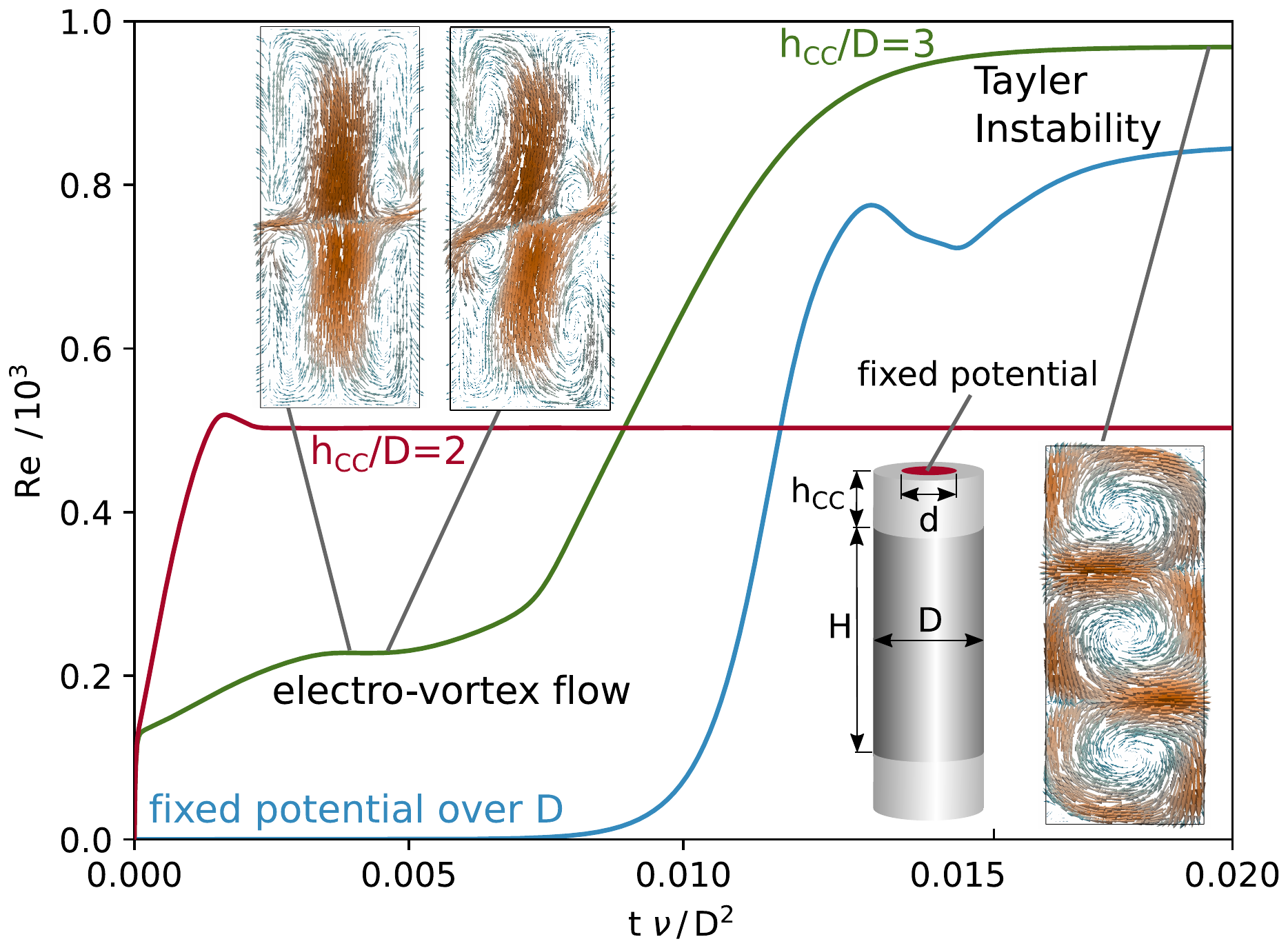}
  \caption{Reynolds number based on the mean velocity in a cylindrical
    liquid metal column vs.~time in viscous units. The applied current
    density corresponds to $Ha = 100$, column diameter and height are
    ($D=1$\,m, $H=2.4$\,m). The material parameters correspond to Na
    at 580\,$\mbox{}^\circ$C. $h_{\text{CC}}$ denotes the height of
    the current collectors, their conductivity is five times that of
    sodium. The centered area with fixed potential has a diameter
    $d = 0.5 D$. Insets show velocity snapshots in a meridional
    plane. Adapted from \cite{WeberGalindoPriedeEtAl:2015}.}
  \label{fig:TI_and_EVF_Weber_et_al}
\end{figure}

Electro-vortex flow almost certainly occurs in typical liquid metal
batteries because current diverges from the negative current collector
to the casing, which serves as positive current collector. Changing
the size and aspect ratio of the current collectors and electrodes can
impede or promote electro-vortex flow in
simulations~\cite{WeberGalindoPriedeEtAl:2015}. 
Because electro-vortex flow is not an instability, there are no
critical dimensionless parameters below which it disappears. Even
small liquid metal batteries supporting gentle currents are
susceptible to finite electro-vortex flow, which may make it a more
important engineering consideration than the Tayler instability. In
fact, an experimental device initially designed to produce the Tayler
instability also drove measurable motion at currents too small for
instability~\cite{StaraceWeberSeilmayerEtAl:2015} after holes for UDV
probes were drilled in the current collectors. The motion stopped
immediately when current was turned off, a behavior inconsistent with
thermal convection; probably electro-vortex flow was the cause. In
fact, electro-vortex flow seems to suppress the Tayler instability in
simulations for thin current collectors (curve $h_{\text{CC}}/D=2$ of Fig.~\ref{fig:TI_and_EVF_Weber_et_al}), though
the Tayler instability, once it sets in, suppresses electro-vortex
flow, see the $h_{\text{CC}}/D=3$ curve and the insets in Fig.~\ref{fig:TI_and_EVF_Weber_et_al}.
Narrow current collectors and thin fluid layers produce current distributions that diverge more and therefore drive stronger electro-vortex flow~\cite{StefaniGalindoKasprzykEtAl:2016}. Intermetallic solids floating at the interface between the positive electrode and electrolyte could also cause electro-vortex flow, since their conductivity is typically lower than the surrounding melt. 

Electro-vortex flow also competes with thermal convection driven by Joule heating. Current diverges from the negative current collector at the top of the battery, so electro-vortex flow tends to form a poloidal roll directed inward at the top and descending along the central axis. Joule heating, on the other hand, tends to cause a poloidal roll that rises along the central axis, where current and therefore heat are concentrated. Both opposing phenomena grow stronger as current increases. In the limiting case of very high current, we would expect thermal effects to dominate, since the rate of Joule heating grows as the square of the current, whereas the work done by electro-vortex flow grows proportionally with current. Still, because both phenomena also depend on the spatial distribution of current, their competition is probably complicated and requires further study. 

Davidson~\cite{Davidson:2001} considered the analogous problem of competition between electro-vortex flow and thermal convection in vacuum arc remelting, in which current diverges from a small electrode through a hemispherical pool of molten metal. The resulting flow can be characterized with the dimensionless quantity 
\[
\chi = \frac{I}{2\pi\kappa} \left( \frac{\mu}{\rho} \right)^{1/2} \left( \frac{3\kappa^2}{g \alpha_T L^3 \Delta T} \right)^{3/7},
\]
where $I$ is the current. When $\chi < 0.4$, thermal convection dominates, and fluid rises along the central axis, as expected. When $\chi > 1.4$, electro-vortex flow dominates, and fluid descends along the central axis. At intermediate values of $\chi$, the two mechanisms have similar strength, and their competition produces more complicated flow patterns involving multiple circulating rolls. Though liquid metal batteries typically involve different sizes, geometries, and materials than remelters, $\chi$ can probably predict the relative importance of thermal convection and electro-vortex flow nonetheless, though the values $\chi < 0.4$ and $\chi > 1.4$ will likely change. However, since $\chi$ is written in terms of a fixed temperature differential $\Delta T$, it may be a better predictor for convection driven by temperature differences at the boundaries than for convection driven by internal Joule heating. 

Like other motions, electro-vortex flow can enhance battery performance by keeping each layer more chemically uniform, but can destroy a battery by rupturing the electrolyte layer. A good predictor of rupture is the Richardson number
\[
Ri = \frac{( \rho_2 / \rho_1 - 1 ) g L}{\left< u \right>^2},
\]
a ratio of the gravitational energy of stratification to the kinetic energy of flow~\cite{HerremanNoreCappaneraGuermond:2015,StefaniGalindoKasprzykEtAl:2016}. Here $\rho_2$ is the density of the electrolyte and $\rho_2$ is the density of the electrode. When $Ri<1$, rupture is likely. 

\section{Summary \& Future Directions}
\label{sec:summary}

In this section, we conclude with a brief summary of the fluid mechanics of liquid metal batteries, especially thermal convection, compositional convection, Marangoni flow, and electro-vortex flow, including a few words about the importance of aspect ratio. Finally, we close with a discussion of open questions and future research directions, including richer experimental measurements, more realistic simulations, applications to battery design, size effects, temperature effects, and the role of solid separators. 

\subsection{Summary}
We have considered the fluid mechanics of liquid metal batteries,
which are made from two liquid metal electrodes and a molten salt
electrolyte, without solid separators. Intended primarily for
grid-scale storage, liquid metal batteries involve similar phenomena
to those in technologies that were invented earlier, including
aluminum smelters, aluminum refinement cells, and thermally
regenerative electrochemical systems. Fluid flow may occur in the
electrolyte or the positive electrode, and affects mass transfer in
both layers, but is likely negligible in the negative electrode,
where a metal foam hinders motion. Flow can destroy liquid metal batteries if it becomes vigorous enough to rupture the electrolyte layer. However, flow can also be beneficial. In parameter regimes common in today's liquid metal batteries, flow is likely important for preventing formation of solid intermetallic phases, which can swell and cause electrical shorts between the positive and negative electrodes, destroying batteries. The extent of swelling depends on the density difference between intermetallic and liquid metal. Intermetallic phases are most likely to form during rapid discharge, but cannot form during charge, so rapid charging poses no danger. 

In current liquid metal battery designs, the primary mechanisms driving flow include thermal convection, compositional convection, Marangoni flow, and electro-vortex flow. Because the resistivity of the electrolyte layer is four orders of magnitude higher than the electrode layers, nearly all Joule heating occurs in the electrolyte, and thermal convection is likely strongest there. The positive electrode, below the electrolyte, is usually subject to stable temperature stratification that may actually hinder flow. Thermal convection in the electrolyte layer is constrained~--- but slowed little or none~--- by the magnetic fields produced by battery currents. The low Prandtl number of the molten salt electrolyte causes its convection to have different properties than a high-$Pr$ fluid like air. 

The magnitude of Marangoni flow in liquid metal batteries is difficult to estimate because the surface tension between molten salts and liquid metals~--- and its variation with temperature and composition~--- has rarely been characterized. That said, because the electrolyte layer is thin, because the layer is subject to intense Joule heating, and because liquid metals have unusually high surface tension in vacuum, Marangoni flow along electrolyte surfaces is probably substantial. 

Electro-vortex flow, driven by an electrical current interacting with its own magnetic field, is present but probably weaker than thermal convection or Marangoni flow in most liquid metal batteries. However, electro-vortex flow may interact with other flow mechanisms can cause important effects. Current can also drive flow by interacting with external magnetic fields, though that effect is also relatively weak. Interface instabilities and the Tayler instability are unlikely in today's liquid metal battery designs, but would become crucial in larger batteries (as discussed further below). The aspect ratio of liquid metal batteries plays an important role. Given two batteries with the same capacity (and therefore same volume), the shallower and broader battery is more prone to interface instabilities, whereas the deeper and narrower battery is more prone to the Tayler instability. Interactions among mechanisms driving flow are also likely, and could trigger instabilities, but little is yet known about the details of their interactions. 

\subsection{Open questions and future directions}
Much remains unknown about the fluid mechanics of liquid metal
batteries, leaving many open research opportunities. First, richer
experimental measurements would substantially advance the field. In
experiments, measuring the flow of an opaque, high-temperature fluid
is unavoidably difficult. Ultrasound Doppler velocimetry (UDV)~\cite{Takeda:1995}
provides richer measurements than most other methods. Each transducer
typically measures one velocity component at a few hundred locations
along the probe's line of sight. Though a line of measurements gives
much more insight into flow shape than point measurements, ultrasound
data is nonetheless sparser than the two- and three-dimensional
measurements commonly available via optical techniques in fluids that
are transparent. (A recent review~\cite{EckertCramerGerbeth:2007} of
velocity measurement methods for liquid metals is also available.)
Experimentalists could contribute substantially to the field by using
multiple probes, either to make single-component measurements along
multiple (carefully chosen)
lines~\cite{StaraceWeberSeilmayerEtAl:2015}, or to make
multi-component measurements along a single line. Recent work
demonstrating two-component, two-dimensional velocity measurement
using phased arrays of ultrasound transducers shows great
promise~\cite{BuettnerNauberBurgerRaebigerFrankeEckertCzarske:2013,RaebigerZhangGalindoFrankeWillersEckert:2014,FrankeRaebigerGalindoZhangEckert:2016,Nauber:2016}. Also, no ultrasound measurements in a
working, three-layer battery have been published. Instead, all
measurements to date were made in a single liquid metal layer, without
electrolyte or a second metal layer~\cite{KelleySadoway:2014,PerezKelley:2015}. However, single-layer experimental models capture only a subset of the fluid mechanics of liquid metal batteries, including Joule heating but not heating via entropy change or heat of formation, which may have significant effects~\cite{Swinkels:1971}; and including thermal convection but not compositional convection. Directly measuring the flow in the positive electrode and the electrolyte of a functioning battery would address many open questions and is a worthy goal for experimentalists. Finally, ultrasound measurements have so far been restricted to temperatures below 230\,$\mbox{}^\circ$C, though the batteries being commercialized are solid at those temperatures. High-temperature ultrasound probes with waveguides have been demonstrated~\cite{EckertGerbethMelnikov:2003} and could be applied to batteries. 

Second, simulation results incorporating more of the relevant physics would substantially advance the field. Simulations to date have sometimes constrained themselves to a single layer~\cite{HerremanNoreCappaneraGuermond:2015} and have often constrained themselves to a subset of possible flow mechanisms, for example, thermal convection~\cite{ShenZikanov:2016}, electro-vortex flow~\cite{WeberGalindoPriedeEtAl:2015}, or the combination of thermal convection and Marangoni flow~\cite{KoellnerBoeckSchumacher:2017}. These simplifications are important first steps to validate codes and build intuition for physical mechanisms at play in liquid metal batteries. Nor are the simulations undertaken trivial~--- low-$Pr$ fluids are tricky and expensive to simulate, and essential material properties have been unavailable. Still, interaction among flow mechanisms are all but certain, so discovering the dynamics of liquid metal batteries and accurately forecasting their behavior will require simulations incorporating multiple mechanisms. Measuring material properties, especially the surface tensions between liquid metals and molten salts, along with their dependence on temperature and composition, would enable better-constrained simulations. Simulating three-layer liquid metal batteries including multiple flow mechanisms would be a great step forward. Among the long list of mechanisms, it seems that thermal convection, compositional convection, Marangoni flow, and electro-vortex flow are most relevant for today's liquid metal batteries. Heat of formation and entropy change may also be significant~\cite{Swinkels:1971}. In simulations of larger batteries, surface instabilities, the Tayler instability, and forces caused by external magnetic fields may have stronger effects. 

Third, richer experimental measurements and simulations incorporating more of the relevant physics would allow improved battery design. Liquid metal batteries involve a large number of design parameters (electrode materials, electrolyte material, size, shape, temperature, current density, etc.) and a large number of engineering metrics (cost, voltage, temperature, cycle life, etc.). So far, materials choices have been considered most carefully, and cost has been identified as the primary engineering metric. Still, trade-offs are common, and the technology is new enough that the long-term consequences of design choices are not always clear. For example, Li$||$PbSb batteries achieve higher voltage and lower cost per kWh than Li$||$Bi batteries~\cite{WangJiangChungOuchiBurkeBoysenKimMueckeSadoway:2014,Ning:2015}, but if Li$||$Bi batteries last much longer because of their ability to eliminate intermetallic growth electrochemically, their lifetime cost may be lower. Improved fluid mechanical experiments and simulations might be accurate enough to predict the rate of intermetallic growth and answer the question of material choice. 

Battery size is a particularly interesting design parameter for fluid
mechanical investigation. Aluminum smelters are typically large, with
horizontal dimensions on the order of 3~m~$\times$~10~m, which lowers
the cost of aluminum. Similarly, liquid metal batteries might provide
storage at lower cost if they were larger. For the same energy (or
power), larger batteries require less container material and less
power electronics, offering substantial cost reductions. However,
larger batteries are more susceptible to electrolyte rupture by
surface instabilities or the Tayler instability (though aspect ratio also plays an important role). Preliminary experiments also show them to be more susceptible to shorts due to intermetallic
growth. Today's prototypes are kept small (200~mm) to avoid those
drawbacks, but improved understanding of the fluid mechanics might
overcome them and allow larger, lower-cost batteries. In aluminum
production, surface instabilities are damped by thickening the
electrolyte to about 4~cm~\cite{Davidson:2000}, at the cost of
requiring more voltage~--- and therefore more energy~--- to drive
current. In batteries, however, that mitigation strategy is
unavailable, since the available voltage is fundamentally limited by
electrode materials, and typically $\lesssim 1$\,V. Improved understanding of the fluid mechanics of liquid metal batteries might reveal alternate strategies. One question of battery size has already been addressed: simulations have shown that batteries designed small enough (and with the right aspect ratio) can avoid the Tayler instability altogether~\cite{HerremanNoreCappaneraGuermond:2015}. Surface instabilities can probably be avoided with a similar design strategy, and simulations (or perhaps experiments) could determine the required size and shape precisely.

Fourth, liquid metal battery fluid mechanics and device designs may
change radically if low-temperature chemistries are
discovered. Battery prototypes being commercialized operate at
475\,$\mbox{}^\circ$C, mainly because the melting point of their molten salt electrolyte is nearly that high. Batteries stacked in large arrays can maintain that temperature by Joule heating if they are used regularly, but when unused, they require extra heat, adding cost. Reducing the operating temperature would reduce that cost and probably have greater effects on other costs. Lower temperatures would cause slower chemical kinetics of side reactions that reduce battery efficiency and corrode containers and current collectors. Lower temperatures would ease the design of seals. Much lower temperatures would allow less expensive container materials, like plastics instead of stainless steel. Thus substantial effort is being dedicated to the search for viable low-temperature liquid metal battery chemistries~\cite{SpatoccoOuchiLambotteBurkeSadoway:2015,SpatoccoSadoway:2015,AshourYinOuchiKelleySadoway:2017,LalauIspasWeierBund:2015,LalauDimitrovaHimmerlichIspasWeierKrischokBund:2016}. If found, they would have significantly different material properties than today's liquid metal batteries and therefore give rise to significantly different fluid mechanics. For example, the lowest-temperature systems use electrolytes composed of room temperature ionic liquids instead of molten salts~\cite{LalauIspasWeierBund:2015,LalauDimitrovaHimmerlichIspasWeierKrischokBund:2016}. Ionic liquids, however, have typical conductivities orders of magnitude smaller than those of molten salts. Liquid metal batteries made with ionic liquids, therefore, would require much thinner electrolyte layers to produce the same voltage, and Joule heating would be still more isolated to the electrolyte layer. The relative magnitude of flow mechanisms (thermal convection, Marangoni flow, electro-vortex flow, instabilities, etc.) may also be different in low-temperature batteries. If low-temperature liquid metal batteries become possible, many practical and interesting questions of fluid mechanics will arise. 

Fifth and finally, though we have focused on liquid metal batteries without solid separators, there are existing technologies~\cite{BonesTeagleBrookerCullen:1989,Sudworth:2001} as well as proposed next-generation batteries~\cite{XuKjosOsenEtAl:2016,XuMartinezOsenKjosKongsteinHaarberg:2017} that include them. The presence of a solid separator radically changes the fluid mechanics and design of batteries. Without any interfaces between two different liquids, neither Marangoni flow nor surface instabilities are possible. Instead, both electrodes are subject to no-slip boundary conditions at the separator. Solid separators eliminate the possibility of rupture due to fluid mechanics and allow electrodes to be positioned side-by-side instead of being stacked vertically. Solid separators have electrical conductivity substantially lower than molten salt~--- about six orders of magnitude lower than liquid metals~--- so heat production would be even more concentrated. High-temperature batteries using solid separators come with their own set of interesting and practical fluid mechanical questions. 

The fluid mechanics of liquid metal batteries is an exciting topic, involving an increasing number of researchers and a large number of open questions. Enabling large-scale storage to make electrical grids more robust while incorporating more wind and solar generation would make a tremendous social impact. Interactions among mass transport, heat transport, multiphase flow, magnetohydrodynamics, and chemical reaction make batteries complicated and interesting. We urge and encourage researchers to focus on problems that are both practical for enabling battery technology and interesting for broadening human knowledge. 

\begin{acknowledgment}
  The authors are grateful to F.\ Stefani for comments on an early
  draft of the manuscript. This work was supported by the National
  Science Foundation under award number CBET-1552182 and by
  Helmholtz-Gemeinschaft Deutscher Forschungszentren (HGF) in frame of
  the Helmholtz Alliance ``Liquid metal technologies''
  (LIMTECH). Fruitful discussions with Valdis Bojarevi{\v c}s, Wietze
  Herreman, Gerrit Horstmann, Caroline Nore, Takanari Ouchi, Donald
  Sadoway, Norbert Weber, and Oleg Zikanov are gratefully
  acknowledged.
\end{acknowledgment}

\begin{nomenclature}
    \entry{$a_{M^{z+}}$}{activity of the cation of M}
    \entry{$a_{M}$}{activity of M}
    \entry{$a_{M(N)}$}{activity of M alloyed with N}
    \entry{$\mathcal{A}_\text{g}$}{interface interaction parameter}
    \entry{$B$}{characteristic magnetic field strength}
    \entry{$B_\varphi$}{azimuthal component of magnetic field}
    \entry{$Bo$}{dynamic Bond number}
    \entry{$X_0$}{characteristic concentration}
    \entry{$D$}{material diffusivity}
    \entry{$E$}{voltage}
    \entry{$E_{\text{OC}}$}{open circuit voltage}
    \entry{$F$}{Faraday constant}
    \entry{$\bm{F}$}{force per unit mass}
    \entry{$\bm{F_L}$}{Lorentz force per unit volume}
    \entry{$G$}{Galileo number}
    \entry{$g$}{acceleration due to gravity}
    \entry{$Ha$}{Hartmann number}
    \entry{$H_C$}{cryolite layer thickness}
    \entry{$H_E$}{aluminum layer thickness}
    \entry{$I$}{current}
    \entry{$I_\text{h}$}{horizontal perturbation current}
    \entry{$J$}{current density}
    \entry{$L$}{vertical thickness}
    \entry{$L_x$}{x-direction size}
    \entry{$L_y$}{y-direction size}
    \entry{$L_z$}{z-direction size}
    \entry{$m$}{azimuthal wave number}
    \entry{M}{alkali or earth alkali metal}
    \entry{$Ma$}{thermal Marangoni number}
    \entry{N}{heavy or half metal}
    \entry{$Nu$}{Nusselt number}
    \entry{$Pm$}{magnetic Prandtl number}
    \entry{$Pr$}{Prandtl number}
    \entry{$Q$}{total heat flux}
    \entry{$R$}{gas constant}
    \entry{$R_{text{E}}$}{ohmic resistance electrolyte}
    \entry{$Ra$}{thermal Rayleigh number}
    \entry{$Ra_\mathrm{crit}$}{critical Rayleigh number for flow onset}
    \entry{$Ra_X$}{compositional Rayleigh number}
    \entry{$Re$}{Reynolds number}
    \entry{$Ri$}{Richardson number}
    \entry{$t$}{time}
    \entry{$T$}{temperature}
    \entry{$t$}{time}
    \entry{$\bm{u}$}{velocity}
    \entry{$u_j$}{velocity component, using indicial notation}
    \entry{$U$}{characteristic flow velocity}
    \entry{$X$}{concentration of negative electrode material}
    \entry{$z$}{valency}
    \entry{$\alpha_T$}{thermal coefficient of volumetric expansion}
    \entry{$\alpha_X$}{solutal coefficient of volumetric expansion}
    \entry{$\beta$}{Sele criterion for metal pad instability}
    \entry{$\beta_\text{cr}$}{Bojarevi{\v c}s-Romero criterion for metal pad instability}
    \entry{$\Delta T$}{characteristic temperature difference}
    \entry{$\Delta X$}{characteristic concentration difference}
    \entry{$\Delta \rho_{\text{CE}}$}{characteristic density difference}
    \entry{$\Delta \sigma$}{characteristic surface tension difference}
    \entry{$\eta^{mn}$}{interface wave amplitude}
    \entry{$\eta_{a, a}$}{activation polarization at the anode}
    \entry{$\eta_{a, c}$}{activation polarization at the cathode}
    \entry{$\eta_{c, a}$}{concentration polarization at the anode}
    \entry{$\eta_{c, c}$}{concentration polarization at the cathode}
    \entry{$\kappa$}{thermal diffusivity}
    \entry{$\mu$}{magnetic permeability}
    \entry{$\nu$}{kinematic viscosity}
    \entry{$\rho$}{density}
    \entry{$\sigma$}{surface tension}
    \entry{$\sigma_E$}{electrical conductivity}
    \entry{$\tau$}{characteristic flow time}
    \entry{$\varphi_{00}$}{standard half-cell potential}
    \entry{$\varphi_{0}$}{half-cell potential}
    \entry{$\omega$}{oscillation frequency}
\end{nomenclature}


\end{document}